\documentclass[11pt]{article}

\usepackage{amsmath,amsthm,amsfonts,amssymb,srcltx}
\usepackage[dvipdfmx]{graphicx}
\usepackage[all]{xy}
\usepackage{color}

\def\IZ{\mathbb {Z}}

\def\IR{\mathbb {R}}
\def\IC{\mathbb {C}}
\def\IP{\mathbb {P}}

\theoremstyle{plain}
  \newtheorem{prob}{Problem}[section]

  \newtheorem{thm}[prob]{Theorem}

    \newtheorem{defi}[prob]{Definition}
\theoremstyle{remark}
  \newtheorem{remark}[prob]{Remark}

\renewcommand{\thefootnote}{\fnsymbol{footnote}}

\makeatletter
 \renewcommand{\theequation}{%
       \thesection.\arabic{equation}}
 \@addtoreset{equation}{section}
\makeatother

\makeatletter
\def\eqnarray{%
 \stepcounter{equation}%
 \let\@currentlabel=\theequation
 \global\@eqnswtrue
 \global\@eqcnt\z@
 \tabskip\@centering
 \let\\=\@eqncr
 $$\halign to \displaywidth\bgroup\@eqnsel\hskip\@centering
 $\displaystyle\tabskip\z@{##}$&\global\@eqcnt\@ne
 \hfil$\displaystyle{{}##{}}$\hfil
 &\global\@eqcnt\tw@$\displaystyle\tabskip\z@{##}$\hfil
 \tabskip\@centering&\llap{##}\tabskip\z@\cr}
\makeatother

\begin{document}
\begin{titlepage}

\begin{center}
\vspace*{1cm}
{\Large \bf
Local B-model Yukawa couplings from A-twisted correlators}
\vskip 1.5cm
{\large Yoshinori Honma${}^{a}$\footnote[2]{yhonma@law.meijigakuin.ac.jp} and 
Masahide Manabe${}^b$\footnote[3]{masahidemanabe@gmail.com}}
\vskip 1.0em
{\it 
${}^a$%
Institute of Physics, Meiji Gakuin University, \\
Yokohama, Kanagawa 244-8539, Japan \\

${}^b$%
Max-Planck-Institut f\"ur Mathematik\\
Vivatsgasse 7, 53111 Bonn, Germany\\}
\end{center}
\vskip2.5cm

\begin{abstract}
Using the exact formula for the A-twisted correlation functions of the 
two dimensional $\mathcal{N}=(2,2)$ 
gauged linear sigma model, we reconsider the computation of the B-model Yukawa couplings of the local toric 
Calabi-Yau varieties. Our analysis is based on an exact result that has been evaluated from the supersymmetric 
localization technique and careful treatment of its application. We provide a detailed description of a procedure to 
investigate the local B-model Yukawa couplings and also test our prescription by comparing the results with known 
expressions evaluated from the local mirror symmetry approach. 
In particular, we find that the ambiguities of classical 
intersection numbers of a certain class of local toric Calabi-Yau varieties discovered previously can be interpreted as 
degrees of freedom of the twisted mass deformations.
\end{abstract}
\end{titlepage}


\renewcommand{\thefootnote}{\arabic{footnote}} \setcounter{footnote}{0}

\section{Introduction}\label{sec:introduction}

The two dimensional  $\mathcal{N}=(2,2)$ gauged linear sigma model (GLSM) 
\cite{Witten:1993yc} has played significant roles in the study of 
topological quantum field theories and mirror symmetry. 
A physical proof of mirror symmetry in \cite{Hori:2000kt} 
makes use of the duality properties of this model. Shortly after the work of \cite{Witten:1993yc}, 
the topological A-twisted version of the GLSM was studied in \cite{Morrison:1994fr} and the genus 
zero correlation functions were exactly computed for the A-models with target space a toric variety or 
a toric Calabi-Yau hypersurface. 
These developments led to a conjecture called 
the toric residue mirror conjecture 
\cite{Batyrev:2002mi,Batyrev:2002fd} (later proven in \cite{Borisov:05,SzVe:2003,Karu:05,SzVe:06}), 
which states the existence of an equality between an integration over the toric compactification\footnote{Toric 
compactifications can be generalized to stable toric quasimaps \cite{CiocanKim:10} allowing 
some degeneracies of the moduli space of maps 
and further to stable quasimaps \cite{CiocanKimMaulik:14}, which also include non-abelian varieties.} 
of the moduli space 
of genus zero maps in the A-model 
and the toric residue \cite{Cox:96} over the space of a holomorphic section in a toric variety in the B-model (see also \cite{Givental:98}). 
The toric residue mirror conjecture provides an efficient method to compute 
the B-model Yukawa couplings 
exactly for the mirror varieties of toric Calabi-Yau complete intersections.

On the other hand, by using the supersymmetric localization technique 
\cite{Pestun:2016zxk}, the exact formula for the A-twisted $\mathcal{N}=(2,2)$ 
GLSM correlation functions on a supersymmetric two-sphere background 
has been clarified in \cite{Closset:2015rna,Benini:2015noa}.\footnote{
It is worth noting that the supersymmetric localization of $\mathcal{N}=(2,2)$ gauge theories has also been performed
on a different two-sphere background, which corresponds to a fusion of the $A$- and $\bar{A}$- twists on two hemispheres \cite{Benini:2012ui,Doroud:2012xw}.} 
This approach does not depend on a particular choice of gauge group and 
enables us to calculate the Yukawa couplings for GLSMs with non-abelian gauge groups efficiently \cite{Closset:2015rna,Ueda:2016wfa} (see also \cite{Kim16}). 
Meanwhile, the application for models with non-compact target spaces has not been thoroughly investigated.
The aim of this work is to fill this gap and establish an explicit formalism to investigate the local B-model Yukawa 
couplings from the A-twisted GLSM correlation functions on local toric Calabi-Yau varieties.

In this paper, we will explicitly demonstrate how to apply the exact formula 
for A-twisted GLSM correlators 
in \cite{Closset:2015rna} for the explicit computation of the B-model Yukawa couplings of local toric Calabi-Yau varieties. 
We remind ourselves that the application of the formula to the local toric Calabi-Yau varieties requires careful treatment of the mass 
parameters of the chiral multiplets.\footnote{We refer the reader to \cite{Gerhardus:2018zwb} where several new aspects 
of twisted mass deformations have been clarified.} 
Then we propose how to determine the proper mass deformations to conduct the exact calculation of the A-twisted GLSM 
correlators and check that the resulting A-twisted correlators coincide with known results for the B-model Yukawa couplings 
evaluated by the local mirror symmetry approach \cite{Chiang:1999tz,Forbes:2005xt,Haghighat:2008gw}. 
To the best of our knowledge, this relationship has not been thoroughly investigated before.
Moreover, as a by-product, we also find that an ambiguity of the classical intersection numbers of a certain class of local toric 
Calabi-Yau varieties argued before in \cite{Forbes:2005xt} can admit a new interpretation.

This paper is organized as follows. First we will take a brief look at the exact results for the GLSM correlation 
functions in Section \ref{sec:glsm}. In Section \ref{sec:local_b_yukawa}, we explain the details of the topological properties of local toric 
Calabi-Yau varieties and propose the rules of mass assignment for applying the formula to non-compact backgrounds. Then in Section 
\ref{sec:examples} we will evaluate the GLSM correlators for several examples of local toric Calabi-Yau threefolds and fourfolds, and demonstrate 
that the resulting expressions are completely consistent with known results obtained by the local mirror symmetry approach. 
Finally we will conclude and propose several future directions. 
In Appendix \ref{sub:I_glsm}, we collect the building blocks of the localization formula, which are useful for understanding the mirror transformation. List of local B-model Yukawa couplings for the local del 
Pezzo surface $K_{dP_2}$ and the local $A_2$ geometry are summarized 
in Appendixes \ref{app:dp2} and \ref{app:a2}, respectively.

\section{$\mathcal{N}=(2,2)$ GLSM on the $\Omega$-deformed two-sphere}\label{sec:glsm}

Here we will take a brief look at the two dimensional $\mathcal{N}=(2,2)$ GLSM on the $\Omega$-deformed 
two-sphere $S^2_{\hbar}$, which is a one-parameter deformation of the A-twisted sphere. As shown 
in \cite{Closset:2015rna}, the supersymmetric localization formula considerably simplifies the exact 
calculations of the A-twisted correlation functions and realizes the quantum cohomology relations appropriately.

\subsection{Exact formula for GLSM correlation functions}\label{subsec:correlator}

Following the approach to curved space rigid supersymmetry advocated in \cite{Festuccia:2011ws},
supersymmetric backgrounds in two dimensions were studied in detail in \cite{Closset:2014pda}. A
remarkable result is that the topological A-twist on the two-sphere admits an interesting $U(1)$ 
equivariant deformation called the $\Omega$-deformation \cite{Nekrasov:2003rj}. This deformation
can be characterized by an equivariant parameter $\hbar$, which corresponds to a non-trivial 
expectation value for the graviphoton field. The ordinary topological A-twist is obtained by setting 
$\hbar =0$.

Let $G$ be a rank $\mathrm{rk}(G)$ gauge group with Lie algebra $\mathfrak{g}$ and Cartan 
subalgebra $\mathfrak{h}\subset \mathfrak{g}$. In \cite{Closset:2015rna}, an exact formula for the 
correlation functions of an $\mathcal{N}=(2,2)$ GLSM with gauge group $G$ on the $\Omega$-deformed two-sphere $S^2_{\hbar}$ was clarified as
\begin{align}
\left< \left. \mathcal{O}_{1} (\mathbf{u}) \right|_N \left. \mathcal{O}_{2} (\mathbf{u}) \right|_S \right>_{\hbar}
=\frac{1}{\left|\mathcal{W}\right|}\sum_{\mathbf{d} \in {\IZ}^{\mathrm{rk}(G)} }\mathbf{z}^{\mathbf{d}}
\sum_{\mathbf{u}_*}
\mathop{\mathrm{JK\textrm{-}Res}}_{\mathbf{u}=\mathbf{u}_*}\left[\mathsf{Q}_*, \eta\right]\,
\mathbf{I}_{\mathbf{d}}\left(\mathcal{O}_{1}\mathcal{O}_{2}\right),
\label{corr formula}
\end{align}
by using the supersymmetric localization technique (see also \cite{Benini:2015noa}). Here 
$\left. \mathcal{O}_{1} (\mathbf{u}) \right|_N$ and $\left. \mathcal{O}_{2} (\mathbf{u}) \right|_S$ 
are gauge invariant operators constructed from the complex scalars 
$\mathbf{u}=\left(u_1,\ldots,u_{\mathrm{rk}(G)}\right) \in \mathfrak{h}\otimes_{\IR}{\IC}$ 
in the vector multiplet inserted at the north and south poles, respectively. 
The order of the Weyl group of $G$ is denoted by $\left|\mathcal{W}\right|$, 
and\footnote{
Here we further assume that the rank of the gauge group $\mathrm{rk}(G)$ 
is equal to the number of central U(1) factors in $G$. 
This condition generically holds 
for the GLSMs with local toric Calabi-Yau backgrounds discussed in the following sections.}
\begin{align}
\mathbf{z}^{\mathbf{d}}=\prod_{i=1}^{r}z_i^{d_i}=\prod_{i=1}^{r} 
\mathrm{e}^{2\pi \sqrt{-1}  \left( \frac{\theta_i}{2\pi}+\sqrt{-1}\xi_i \right) d_i}
\qquad
(i=1,\ldots r),
\label{theFI}
\end{align}
where $\theta_i$ and $\xi_i$ are theta angles and Fayet-Iliopoulos parameters for the central $U(1)^{r}\subset G$. 
The parameters ${\mathbf{d}} =\left(d_1, \ldots, d_{r} \right) $ represent magnetic charges 
for the central $U(1)^{r}$ called GNO charges \cite{Goddard:1976qe}.

The factor $\mathbf{I}_{\mathbf{d}}\left(\mathcal{O}_{1}\mathcal{O}_{2}\right)$ is a differential form given by
\begin{align}
\begin{split}
\mathbf{I}_{\mathbf{d}}\left(\mathcal{O}_{1}\mathcal{O}_{2}\right)&=
\mathcal{O}_{1}\Big(\mathbf{u}-\frac{\mathbf{d}}{2}\hbar\Big)\,
\mathcal{O}_{2}\Big(\mathbf{u}+\frac{\mathbf{d}}{2}\hbar\Big)
Z_{\mathbf{d}}^{\textrm{vec}}(\mathbf{u};\hbar)
\prod_a Z_{\mathbf{d}}^{\Phi_a}(\mathbf{u};\hbar)\,
du_1 \wedge \cdots \wedge du_{\mathrm{rk}(G)},
\label{col_i_oo}
\end{split}
\end{align}
which consists of the one-loop determinants
\begin{align}
Z_{\mathbf{d}}^{\textrm{vec}}(\mathbf{u};\hbar)=
(-1)^{\sum_{\alpha\in\Delta_+}(\alpha(\mathbf{d})+1)}
\prod_{\alpha\in\Delta_+}\Big(\alpha(\mathbf{u})^2-\frac{\alpha(\mathbf{d})^2}{4}\hbar^2\Big),
\label{vec_formula}
\end{align}
for the vector multiplet and
\begin{align}
Z_{\mathbf{d}}^{\Phi_a}(\mathbf{u};\hbar)=(-1)^{\delta_{r_a,2}}
\prod_{\rho_a\in R_a}\hbar^{r_a-\rho_a(\mathbf{d})-1}\,
\frac{\Gamma\Big(\frac{\rho_a(\mathbf{u})+\lambda_a}{\hbar}+\frac{r_a-\rho_a(\mathbf{d})}{2}\Big)}{\Gamma\Big(\frac{\rho_a(\mathbf{u})+\lambda_a}{\hbar}-\frac{r_a-\rho_a(\mathbf{d})}{2}+1\Big)},
\label{mat_formula}
\end{align}
for the chiral matter multiplets $\Phi_a$ in a representation $R_a$ with $R$-charge $r_a$ and the twisted mass $\lambda_a$.
Here $\Delta_+$ is the set of positive roots and $\rho_a$ denote the weights of $R_a$. The products $\alpha(*)$ and $\rho_a(*)$ 
are defined by the canonical pairing. 

The one-loop determinant for chiral multiplets $\eqref{mat_formula}$ has poles along hyperplanes
defined from the $\Gamma$-function in the denominators. For later convenience, we will use a collective  
form for the gauge charges as $\mathsf{Q}=\left\{Q_i \in {\IZ}^{\mathrm{rk}(G)}\subset \mathfrak{h}^*\right\}$ where 
$i$ labels all the components of the multiplets of the GLSM. The hyperplanes can intersect at a point 
$\mathbf{u}=\mathbf{u}_*=(u_1^*,\ldots,u_{\mathrm{rk}G}^*)$ and realize a codimension $\mathrm{rk}(G)$ pole. 
Such intersecting hyperplanes simultaneously specify a subset of charge vectors with at least $\mathrm{rk}(G)$ elements 
and we denote it by $\mathsf{Q}_* \subset \mathsf{Q}$.

A crucial ingredient of the exact formula (\ref{corr formula}), whose treatment will be described in the next section, is 
the Jeffrey-Kirwan residue operation \cite{JeKi:1993}\footnote{See also \cite{Witten:1992xu,BrVer:1999} for the early 
developments.} $\mathrm{JK\textrm{-}Res}\left[\mathsf{Q}_*, \eta \right]$ 
at $\mathbf{u}=\mathbf{u}_*$ 
depending on a choice of a covector $\eta \in \mathfrak{h}^*$. 
As adopted in \cite{Closset:2015rna}, we always choose $\eta$ to be parallel and pointing 
in the same direction as $\xi_i$ and therefore $\eta$ specifies the phase of the model.

\subsection{The Jeffrey-Kirwan residue operation}\label{subsec:JK_res}
 
In this section, we will briefly review the Jeffrey-Kirwan residue operation. For more details, we refer the reader 
to \cite{SzVe:2003, Benini:2013xpa}. 

\begin{defi}\label{def:proj_hyp}
When $\mathsf{Q}_* \subset \mathsf{Q}$ lies within an open half-space of $\mathfrak{h}^*$, the associated intersection point 
$\mathbf{u}=\mathbf{u}_*$ is called the projective point. Such an arrangement of hyperplanes is called a projective arrangement.
\end{defi}

\begin{defi}
Consider an integrand $\frac{du_1 \wedge \cdots \wedge du_{\mathrm{rk}(G)}} 
{Q_1(\mathbf{u})\cdots Q_{\mathrm{rk}(G)}(\mathbf{u})}$ with a projective point $\mathbf{u}_*=\mathbf{0}$ 
and $\mathsf{Q}_*=\{Q_1,\ldots,Q_{\mathrm{rk}(G)}\}$. 
The cases with generic $\mathbf{u}_*$ can be realized by shifting the coordinates appropriately.
Then the Jeffrey-Kirwan residue 
$\mathop{\mathrm{JK\textrm{-}Res}}_{\mathbf{u}=\mathbf{u}_*}\left[\mathsf{Q}_*, \eta\right]$ is given by
\begin{align}
\mathop{\mathrm{JK\textrm{-}Res}}_{\mathbf{u}=\mathbf{0}}\left[\mathsf{Q}_*,\eta\right]
\frac{du_1 \wedge \cdots \wedge du_{\mathrm{rk}(G)}}
{Q_1(\mathbf{u})\cdots Q_{\mathrm{rk}(G)}(\mathbf{u})}=
\begin{cases}
\frac{1}{\left|\det(Q_1,\ldots,Q_{\mathrm{rk}(G)})\right|}\ \ 
&\textrm{if}\ \eta \in \mathrm{Cone}(Q_1,\ldots,Q_{\mathrm{rk}(G)}),
\\
0
&\textrm{if}\ \eta \notin \mathrm{Cone}(Q_1,\ldots,Q_{\mathrm{rk}(G)}),
\end{cases}
\nonumber
\end{align}
where $\mathrm{Cone}(Q_1,\ldots,Q_{\mathrm{rk}(G)})$ is the closed cone spanned by $Q_1,\ldots,Q_{\mathrm{rk}(G)}$.
\end{defi}

A more constructive definition can be  represented as follows. 
Let $\mathcal{F}\mathcal{L}(\mathsf{Q}_*)$ be 
a finite set of flags
$$
F=\left[F_0=\left\{0\right\} \subset F_1 \subset \cdots \subset F_{\mathrm{rk}(G)}=\mathfrak{h}^*\right],\quad
\dim F_j=j,
$$
such that $\mathsf{Q}_*$ contains a basis of each of the flags $F_j$, $j=1, \ldots, \mathrm{rk}(G)$. 
Then one can choose an ordered set 
$\mathsf{Q}_*^F=(Q_{i_1},\ldots,Q_{i_{{\mathrm{rk}(G)}}})$ 
such that the first $j$ elements $\{Q_{i_m}\}_{m=1}^j$ give a basis of $F_j$.

\begin{defi}
For the above-defined ordered basis $\mathsf{Q}_*^F$ of a flag $F\in \mathcal{F}\mathcal{L}(\mathsf{Q}_*)$, 
the iterated residue $\mathop{\mathrm{Res}}_F$ of 
$\omega=\omega_{1,\ldots, {\mathrm{rk}(G)}}
d Q_{i_1}(\mathbf{u})\wedge \cdots \wedge dQ_{i_{{\mathrm{rk}(G)}}}(\mathbf{u})$ is defined by
\begin{align}
\mathop{\mathrm{Res}}_F \omega
=\mathop{\mathrm{Res}}_{Q_{i_{{\mathrm{rk}(G)}}}(\mathbf{u})=0}\cdots
\mathop{\mathrm{Res}}_{Q_{i_1}(\mathbf{u})=0}\omega_{1,\ldots, {\mathrm{rk}(G)}}.
\nonumber
\end{align}
Here in each step of the residue operations on the right hand side, the higher variables remain to 
be free parameters. Note that 
the iterated residue only depends on the flag $F$, and does not depend on the choice of the ordered basis.
\end{defi}

Let us define the closed simplicial cone for a flag $F\in \mathcal{F}\mathcal{L}(\mathsf{Q}_*)$ as
\begin{align}
\mathfrak{s}^{+}(F, \mathsf{Q}_*)=\sum_{j=1}^{\mathrm{rk}(G)}{\IR}_{\ge 0}\kappa_j^F,\qquad
\kappa_j^F:=\sum_{Q_i\in F_j}Q_i,\quad j=1,\ldots, \mathrm{rk}(G),
\label{def_skappa}
\end{align}
and denote by $\mathcal{F}\mathcal{L}^{+}(\mathsf{Q}_*, \eta)$ a set of flags such that the corresponding 
cone $\mathfrak{s}^{+}(F, \mathsf{Q}_*)$ contains $\eta$. 

\begin{defi}
By the partial sums of the elements of $\mathsf{Q}_*=\{Q_1,\ldots,Q_{n}\}$, 
$n\ge \mathrm{rk}(G)$, define a set
\begin{align}
\Sigma \mathsf{Q}_*=\left\{\sum_{i\in \pi}Q_i,\ \pi \subset \left\{1,\ldots, n \right\} \right\}.
\nonumber
\end{align}
An element $v \in \mathrm{Cone}(\mathsf{Q}_*)$ is called regular with respect to $\Sigma \mathsf{Q}_*$ 
if $v \not\in \mathrm{Cone}_{sing}(\Sigma \mathsf{Q}_*)$, where $\mathrm{Cone}_{\mathrm{sing}}(\Sigma \mathsf{Q}_*)$ 
is the union of the closed cones spanned by $\mathrm{rk}(G)-1$ independent elements of $\Sigma \mathsf{Q}_*$.
A regular $v \not\in \mathrm{Cone}_{sing}(\Sigma \mathsf{Q}_*)$ identifies one chamber in $\mathfrak{h}^*$.
\end{defi}

Then the following theorem can be proved \cite{SzVe:2003}.
\begin{thm}
\label{thm:jk_iterate}
If $\eta \in \mathrm{Cone}(\mathsf{Q}_*)$ is regular with respect to $\Sigma \mathsf{Q}_*$, the Jeffrey-Kirwan residue 
at a projective point $\mathbf{u}=\mathbf{u_*}=\mathbf{0}$ can be written in terms of the iterated residue as
\begin{align}
\mathop{\mathrm{JK\textrm{-}Res}}_{\mathbf{u}=\mathbf{0}}\left[\mathsf{Q}_*,\eta\right]=\sum_{F\in \mathcal{F}\mathcal{L}^{+}(\mathsf{Q}_*, \eta)}
\nu(F)\mathop{\mathrm{Res}}_F,
\label{jk_it_res_thm}
\end{align}
where $\nu(F)=0$ if $\kappa_j^F$, $j=1,\ldots, \mathrm{rk}(G)$ are linearly dependent, 
and $\nu(F)=1$ (resp. $-1$) if $\kappa_j^F$ are linearly independent and the ordered basis $\kappa^F:=(\kappa_1^F,\ldots,\kappa_{\mathrm{rk}(G)}^F)\in \mathfrak{h}^*$ is positively (resp. negatively) oriented, i.e. $\mathrm{sign}(\det \kappa^F)=1$ (resp. $-1$).
\end{thm}

\section{GLSM correlators and local B-model Yukawa couplings}\label{sec:local_b_yukawa}

It has been clarified in \cite{Closset:2015rna} that the A-twisted correlators 
given by \eqref{corr formula} with vanishing $\Omega$-deformation 
precisely give the B-model Yukawa couplings of the mirror 
of compact Calabi-Yau manifolds and non-compact orbifolds.\footnote{See also \cite{Ueda:2016wfa,Kim16} for the
treatment of manifolds with non-abelian GLSM descriptions.}
Here, we consider local toric Calabi-Yau varieties. 
In order to perform the Jeffrey-Kirwan residue operation for these backgrounds appropriately, it is necessary to introduce 
the twisted masses to matter fields and realize a projective hyperplane arrangement of Definition \ref{def:proj_hyp}. 
The main purpose of this paper is to provide a detailed description of this procedure to investigate the local B-model 
Yukawa couplings.

\subsection{Local toric Calabi-Yau varieties and correlation functions}\label{subsec:corr_local_cy}

Consider a local toric Calabi-Yau $m$-fold $X$ described by a symplectic quotient $X={\IC}^n/\!/\!_{\xi}\left({\IC}^*\right)^r$ 
with $m=n-r, \ m \ge 3$.\footnote{We refer the reader to \cite{Cox:2000vi} for a detailed introduction on this subject.} In terms
of the GLSM, this background can be specified by $n$ chiral matter multiplets $\Phi_i$, $i=1,\ldots,n$ with $R$-charge $0$ and 
$U(1)^r$ charge vectors $Q_i \in {\IZ}^r$ satisfying the Calabi-Yau condition $\sum_{i=1}^nQ_i=\mathbf{0}$. One can also assign 
the twisted masses $-\lambda_i$ for the chiral multiplets $\Phi_i$. 
Then a model can be specified by a set of charge vectors and twisted masses as
\begin{align}
\left(
\begin{array}{cccc}
Q_1&Q_2&\cdots&Q_n\\ \hline
\lambda_1&\lambda_2&\cdots&\lambda_n
\end{array}
\right).
\label{glsm_charge}
\end{align}

In the limit of the vanishing $\Omega$-deformation $\hbar=0$, 
the exact formula for correlation functions \eqref{corr formula} 
in the geometric phase reduces to a simple expression 
\begin{align}
\left<\mathcal{O}(\mathbf{u})\right>_{\hbar=0}
=\sum_{\mathbf{d}\in \left({\IZ}_{\ge 0}\right)^r}\mathbf{z}^{\mathbf{d}}
\sum_{\mathbf{u}_*}
\mathop{\mathrm{JK\textrm{-}Res}}_{\mathbf{u}=\mathbf{u}_*}\left[\mathsf{Q}_*,\eta\right]
\mathcal{O}(\mathbf{u})
\prod_{i=1}^n Z_{\mathbf{d}}^{\Phi_i}(\mathbf{u})\,
du_1 \wedge \cdots \wedge du_{r},
\label{lc_yukawa_formula}
\end{align}
where the moduli parameters ${\mathbf{z}} =\left(z_1, \ldots, z_r \right) $ are defined in (\ref{theFI}) 
and the one-loop determinant for chiral multiplets in \eqref{mat_formula} or \eqref{phi_c_loop} takes a reduced form:
\begin{align}
Z_{\mathbf{d}}^{\Phi_i}(\mathbf{u}):=Z_{\mathbf{d}}^{\Phi_i}(\mathbf{u};0)=
\left(Q_i(\mathbf{u})-\lambda_i\right)^{-Q_i(\mathbf{d})-1}.
\end{align}
Choosing the monomials of $\mathbf{u}$ as the operators $\mathcal{O}(\mathbf{u})$, 
we focus on the following GLSM correlators:
\begin{align}
Y_{z_{i_1}\cdots z_{i_m}}(\mathbf{z}):=\left<u_{i_1}\cdots u_{i_m}\right>_{\hbar=0},\quad 
1\le i_1\le \cdots \le i_m \le r.
\label{lc_yukawa_mon}
\end{align}

\begin{remark}
The phase structure of the Calabi-Yau variety $X$ is described by the secondary fan, which consists of 
a set of the charge vectors $\{Q_1,\ldots,Q_n\}$ in $\mathfrak{h}^*$ (see e.g. \cite{Cox:2000vi}).
Although we will consider a particular choice of $\eta \in \mathfrak{h}^*$ inside the geometric phase of 
$X$ for actual computation of the GLSM correlators \eqref{lc_yukawa_mon}, the result does not depend 
on the choice of $\eta$ as long as the Jeffrey-Kirwan residue operation is carried out appropriately.
\end{remark}

\subsection{Criteria for the twisted mass deformations}\label{subsec:rule_mass}

Computations of the GLSM correlators \eqref{lc_yukawa_mon} for local toric Calabi-Yau varieties require careful 
treatment. This is because generically the associated hyperplane arrangement without twisted mass parameters 
$\lambda_i$ becomes non-projective and one should not use the Jeffrey-Kirwan residue operation directly.
Therefore it is necessary to introduce twisted masses for chiral multiplets in a proper way, such that the 
hyperplane arrangement of the model becomes projective.\footnote{Note that the relevance of twisted mass 
deformations has also been clarified in \cite{Gerhardus:2018zwb}
from the other perspective.} Although the necessity of this kind of prescription 
has been argued before, to the authors' knowledge, there has been no satisfactory attempt to
clarify the proper method for the twisted mass insertions. 
Here we propose the following rules as a proper determination of the twisted mass deformations:
\begin{itemize}
\item[{\bf 1.}]
The twisted masses should not be inserted for chiral multiplets with $Q_i \ne \mathbf{0}$ describing a non-compact 
fiber coordinate, or a blow-up coordinate of a singularity.

\item[{\bf 2.}]
Suppose a GLSM of a local toric Calabi-Yau variety of interest has a 
chiral multiplet $\Phi_0$ which is neutral under the $j$-th $U(1)$ gauge symmetry and describes a non-compact fiber coordinate $X_0$ 
by its scalar component.\footnote{A typical example with this property will be shown in (\ref{02P}).}
Let $\lambda_i$, $i=1,\ldots,s$ be twisted masses 
for other chiral multiplets $\Phi_i$ with non-zero charges 
with respect to the $j$-th $U(1)$ gauge symmetry. If a divisor defined by 
$X_0=0$ contains a blow-up coordinate of a singularity, one needs to impose
$$
\lambda_0+\sum_{i=1}^s\lambda_i=0,
$$
for the twisted mass $\lambda_0$ of $\Phi_0$ as a ``Calabi-Yau condition on the divisor''.

\item[{\bf 3.}]
As long as the above requirements are fulfilled, one can turn on generic twisted masses for the
remaining chiral multiplets while respecting the symmetries of the model such as the permutation
of the homogeneous coordinates of the toric variety.

\end{itemize}
In Section \ref{sec:examples}, by using the above prescription for various examples, we will explicitly check 
that the GLSM correlators \eqref{lc_yukawa_mon} give the same results predicted by the mirror symmetry approach. 
Moreover, we find that the ambiguities of the intersection numbers for a certain class of varieties argued 
in \cite{Forbes:2005xt} can be reinterpreted as degrees of freedom of the proper twisted mass deformations.

\begin{remark}
In this paper, we do not consider the so-called non-nef toric varieties such as $\mathcal{O}(k)\oplus\mathcal{O}(-2-k)\to {\IP}^1$ 
with $k \ge 1$ \cite{Forbes:2006sj,Forbes:2006ab,Forbes:2007cy}. It would be interesting to investigate such varieties 
and try to extend our prescription.
\end{remark}

\subsection{Local Yukawa couplings and the mirror map}\label{subsec:local_yukawa}

The topological B-model $m$-point Yukawa couplings of a Calabi-Yau $m$-fold $X^{\vee}$ are defined by 
\begin{equation}
Y_{z_{i_1}\cdots z_{i_m}}(\mathbf{z})=
\int_{X^{\vee}}\Omega(\mathbf{z})\wedge \nabla_{z_{i_1}\partial_{z_{i_1}}}\cdots
\nabla_{z_{i_m}\partial_{z_{i_m}}}\Omega(\mathbf{z})
\ \in\ \textrm{Sym}^m\left(T^*\mathcal{M}\right)\otimes \mathcal{L}^{-2},
\label{local_b_yukawa}
\end{equation}
where $\nabla$ is a flat connection called a Gauss-Manin connection (see e.g. \cite{Cox:2000vi}) and $T^*\mathcal{M}$ is the cotangent bundle of the 
complex structure moduli space $\mathcal{M}$ of the Calabi-Yau $m$-fold $X^{\vee}$. 
$\mathcal{L}$ is a holomorphic line bundle over $\mathcal{M}$ whose section is given by 
the nowhere-vanishing holomorphic $m$-form $\Omega(\mathbf{z})$ on $X^{\vee}$. 
The topological A-model $m$-point Yukawa couplings for the mirror $X$ of $X^{\vee}$ 
can be obtained from \eqref{local_b_yukawa} by changing 
the complex structure moduli parameters $\{\log z_i\}$ into flat coordinates $\{\log q_i\}$ parametrizing $h_i \in H^{1,1}(X)$ as 
\begin{equation}
\widetilde{Y}_{h_{i_1}\cdots h_{i_m}}(\mathbf{q})=\frac{1}{X_0(\mathbf{z}(\mathbf{q}))^2}
\sum_{j_1,\ldots,j_m=1}^m
Y_{z_{j_1}\cdots z_{j_m}}(\mathbf{z}(\mathbf{q}))\,
\frac{\partial \log z_{j_1}(\mathbf{q})}{\partial \log q_{i_1}}\cdots
\frac{\partial \log z_{j_m}(\mathbf{q})}{\partial \log q_{i_m}}.
\label{local_a_yukawa}
\end{equation}
This transformation map is called a mirror map. 
Note that the quantity $X_0(\mathbf{z})^2$ is a square of the monodromy invariant fundamental period of 
$X^{\vee}$ and is equal to 1 for local toric Calabi-Yau varieties, while for compact Calabi-Yau $m$-folds it is a non-trivial function of the moduli parameters $\mathbf{z}$.

Generically the A-model three-point Yukawa coupling 
of a Calabi-Yau threefold, say $X_3$, takes the following form \cite{Candelas:1990rm,Aspinwall:1991ce,
Hosono:1993qy,Hosono:1994ax,Chiang:1999tz}:
\begin{align}
\widetilde{Y}_{h_{i}h_{j}h_{k}}(\mathbf{q})=\kappa_{h_ih_jh_k}+
\sum_{\mathbf{d} \in H_2(X_3,{\IZ})\backslash \{\mathbf{0}\}}
n_{\mathbf{d}}\, \frac{d_i d_jd_k\mathbf{q}^{\mathbf{d}}}{1-\mathbf{q}^{\mathbf{d}}},\qquad
\mathbf{q}^{\mathbf{d}}=q_1^{d_1}q_2^{d_2}\cdots.
\label{a_yukawa_3}
\end{align}
This provides a generating function of integer invariants $n_{\mathbf{d}}$ 
which gives the Gromov-Witten invariants and enumerates 
the number of holomorphic maps $\phi: {\IP}^1 \to X_3$ of class $\mathbf{d} \in H_2(X_3,{\IZ})$ intersecting with  
cycles dual to $h_i$, $h_{j}$, and $h_{k}$. 
Here $\kappa_{h_ih_jh_k}$ is called the classical triple intersection number 
and the summation with respect to $\mathbf{d}$ is taken only over non-negative elements. 

For Calabi-Yau $m$-folds, the $m$-point Yukawa couplings can be factorized into three-point Yukawa couplings in accordance 
with the fusion rules of the underlying Frobenius algebra (see \cite{Greene:1993vm,Mayr:1996sh,Klemm:1996ts} for details). 
The so-obtained A-model three-point Yukawa couplings similarly provide generating functions of genus zero Gromov-Witten 
invariants for $m$-folds. In the case of a Calabi-Yau fourfold $X_4$, the A-model four-point Yukawa couplings are factorized 
into three-point Yukawa couplings as
\begin{align}
\widetilde{Y}_{h_{i}h_{j}h_{k}h_{l}}(\mathbf{q})=\eta^{\alpha\beta}\,
\widetilde{Y}_{H_{\alpha}h_{i}h_{j}}(\mathbf{q})\,
\widetilde{Y}_{H_{\beta}h_{k}h_{l}}(\mathbf{q}),
\label{factorize:yukawa_4}
\end{align}
where $H_{\alpha},H_{\beta} \in H^{2,2}(X_4)$ represent the so-called primary elements generated by the wedge product of 
the elements in $H^{1,1}(X_4)$ and $\eta^{\alpha\beta}$ is the inverse matrix of the intersection matrix associated with 
$H_{\alpha}$ and $H_{\beta}$. The A-model three-point Yukawa couplings $\widetilde{Y}_{h_{i}h_{j}H}(\mathbf{q})$ have
a generic form \cite{Greene:1993vm,Mayr:1996sh,Klemm:1996ts,Klemm:2007in}:
\begin{align}
\widetilde{Y}_{h_{i}h_{j}H}(\mathbf{q})=\kappa_{h_ih_jH}+
\sum_{\mathbf{d} \in H_2(X_4,{\IZ})\backslash \{\mathbf{0}\}}
n_{\mathbf{d}}(H)\, \frac{d_id_j\mathbf{q}^{\mathbf{d}}}{1-\mathbf{q}^{\mathbf{d}}},\qquad
\mathbf{q}^{\mathbf{d}}=q_1^{d_1}q_2^{d_2}\cdots,
\label{a_yukawa_4}
\end{align}
where $\kappa_{h_ih_jH}$ is the classical intersection number associated with cycles dual to $h_i$, $h_{j}$ and $H$.

\subsubsection{Mirror map from the localization formula}

Starting from the factors (\ref{vec_formula}) and (\ref{mat_formula}) in the localization formula, one can find 
building blocks of the Givental $I$-function \cite{Givental:1995,Givental,Coates:2001ewh} from which the mirror
map can be properly derived. Here we go straight to the point; some technical details are relegated to  
Appendix \ref{sub:I_glsm}. 

For a local toric Calabi-Yau $m$-fold $X$ with GLSM description \eqref{glsm_charge}, one can 
construct the Givental $I$-function for $X$  with the following form:
\begin{align}
I_{X}^{\{\lambda_i\}}(\mathbf{z};\mathbf{x};\hbar)=
\mathbf{z}^{\mathbf{x}/\hbar}\sum_{\mathbf{d}\in \left({\IZ}_{\ge 0}\right)^r}
\left(\prod_{i=1}^{n} I_{\mathbf{d}}^{\Phi_i}(\mathbf{x},\lambda_i;\hbar)\right)
\mathbf{z}^{\mathbf{d}},\qquad
\mathbf{z}^{\mathbf{x}/\hbar}=\prod_{j=1}^r z_j^{x_j/\hbar},
\end{align}
which consists of the building blocks defined in \eqref{build_mat0}.
To find the mirror map, it is sufficient to consider the case without twisted mass deformations. 
Expanding the $I$-function with $\lambda_i=0$ around $\hbar = \infty$, we obtain
\begin{align}
I_{X}^{\{\mathbf{0}\}}(\mathbf{z};\mathbf{x};\hbar)=
1+\left(I_{1,1}(\mathbf{z}) x_1+\cdots +I_{1,r}(\mathbf{z}) x_r \right)\hbar^{-1}+O(\hbar^{-2}).
\label{I_expand}
\end{align}
Then the inverse mirror map $\log q_i$, $i=1,\ldots,r$ can be obtained by
$\log q_i = I_{1,i}(\mathbf{z})$ \cite{Chiang:1999tz,Lerche:2001cw} and the result is
\begin{align}
\log \mathbf{q}=\log \mathbf{z}-
\sum_{\mathbf{d}\in \left({\IZ}_{\ge 0}\right)^r \backslash \{\mathbf{0}\}}
{\sum_{I}}'\,
(-1)^{Q_I(\mathbf{d})}\,Q_I\,\frac{(-Q_I(\mathbf{d})-1)!}
{\prod_{i\ne I} Q_i(\mathbf{d})!}\,\mathbf{z}^{\mathbf{d}},
\label{mirror_map}
\end{align}
where we put $1/n!=0$ for $n<0$ and ${\sum}'_{I}$ means that the summation is taken over all $I$ satisfying $Q_I(\mathbf{d})<0$.

\begin{remark}
For a compact Calabi-Yau manifold, the $I$-function 
is generically annihilated by the Picard-Fuchs equation for 
the mirror.\footnote{See \cite{Gerhardus:2018zwb} for a recent development.} 
Then the higher-order terms of the expansion in 
\eqref{I_expand} are expected to 
possess information about the genus zero Gromov-Witten invariants 
(see e.g. \cite{PopaZinger_co}). 
But for a local toric Calabi-Yau variety, say $X_{local}$, the information obtained from the $I$-function without 
twisted mass parameters (i.e. $\lambda_i=0$ for all $i$) is sometimes not enough to get Gromov-Witten invariants. 
For instance, if $\dim H_4(X_{local},{\IZ})=0$, the higher-order terms $O(\hbar^{-2})$ in \eqref{I_expand} vanish and 
cannot say anything about the Gromov-Witten invariants.

It is worth noting that by using the $I$-function with twisted mass parameters 
and the Birkhoff factorization \cite{Coates:2001ewh}, 
the equivariant local mirror symmetry for local toric Calabi-Yau threefolds 
was developed in \cite{Forbes:2006sj,Forbes:2006ab,Forbes:2007cy} and 
the Gromov-Witten invariants have been computed.
\end{remark}

\section{Examples}\label{sec:examples}

Based on the criteria for the twisted mass deformations mentioned in Section \ref{subsec:rule_mass}, 
here we will compute the A-twisted GLSM correlators \eqref{lc_yukawa_mon} for various local nef toric 
Calabi-Yau varieties in the geometric phase and demonstrate that the corresponding local B-model 
Yukawa couplings can be calculated appropriately. A novel aspect of the previously observed ambiguities 
for a certain class of varieties \cite{Forbes:2005xt} will also be discussed.

\subsection{Local toric Calabi-Yau threefolds}\label{subsec:local_cy3}

First we will focus on the local nef toric Calabi-Yau threefolds whose exact properties have been studied by
using the local mirror symmetry \cite{Chiang:1999tz,Forbes:2005xt,Forbes:2006sj,Forbes:2006ab,Forbes:2007cy,Konishi:2010eur}. 
We confirm that the exact formula for the A-twisted GLSM correlation functions matches the previous 
results of the local B-model Yukawa couplings with the aid of our prescriptions. We also see that the 
ambiguities of the intersection numbers in \cite{Forbes:2005xt} can be interpreted as degrees of freedom 
of the twisted mass deformations.

\subsubsection{Resolved conifold: $\mathcal{O}(-1)\oplus\mathcal{O}(-1)\to {\IP}^1$}

\begin{figure}[t]
\centering
\includegraphics[width=150mm]{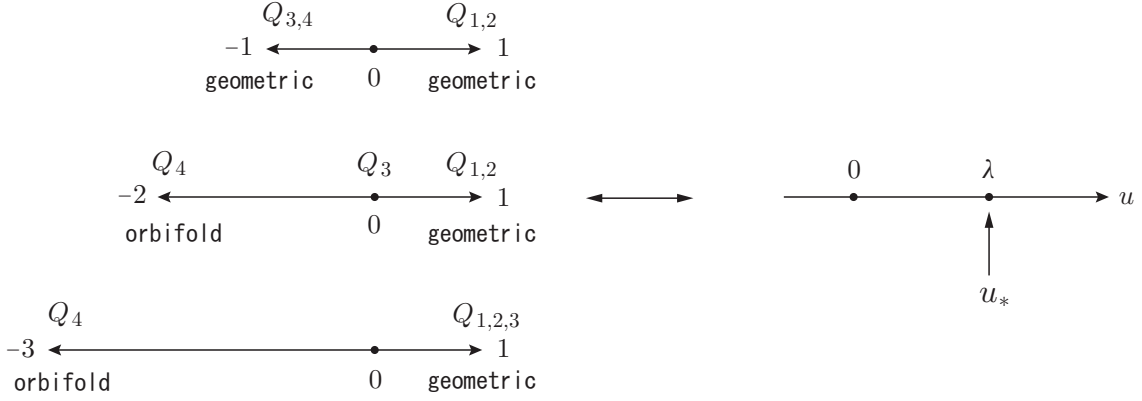}
\caption{\textbf{1)} The upper left diagram describes the secondary fan of the resolved conifold 
and the right one represents the corresponding hyperplanes $\{u=\lambda, u=0\}$ corresponding 
to the charge vectors $\{Q_{1,2}=1, Q_{3,4}=-1\}$. The point $u_*=\lambda$ is a projective point 
associated with $\mathsf{Q}_*=Q_{1,2}$.
\textbf{2)} The middle (resp. lower) left diagram describes the secondary fan of 
$\mathcal{O}(0)\oplus\mathcal{O}(-2)\to {\IP}^1$ (resp. $\mathcal{O}(-3)\to {\IP}^2$) and the right figure again 
describes the hyperplanes $\{u=\lambda, u=0\}$ associated with $\{Q_{1,2}=1, Q_{4}=-2\}$ 
(resp. $\{Q_{1,2,3}=1, Q_{4}=-3\}$). The point $u_*=\lambda$ is a projective point associated with 
$\mathsf{Q}_*=Q_{1,2}$ (resp. $\mathsf{Q}_*=Q_{1,2,3}$). The orbifold phase of the middle (resp. lower) left 
diagram is given by the orbifold geometry ${\IC}\times {\IC}^2/{\IZ}_2$ (resp. ${\IC}^3/{\IZ}_3$).}
\label{fig:local_1}
\end{figure}

Let us consider the resolved conifold described by a $U(1)$ GLSM with the following charge vectors and the 
twisted masses:\footnote{For the resolved conifold as well as the local ${\IP}^2$, several aspects of twisted 
mass deformations have also been discussed in the context of correlator relations in \cite{Gerhardus:2018zwb}.}
\begin{align}
\left(
\begin{array}{cccc}
Q_1&Q_2&Q_3&Q_4\\ \hline
\lambda_1&\lambda_2&\lambda_3&\lambda_4
\end{array}
\right)
=
\left(
\begin{array}{cccc}
1&1&-1&-1\\ \hline
\lambda&\lambda&0&0
\end{array}
\right),
\label{res_coni_charge}
\end{align}
where the inclusion of the twisted mass parameters $\lambda\ne 0$ has been determined in accordance
with the rules in Section \ref{subsec:rule_mass}, 
such that the pole $u_*=\lambda$ of the exact formula associated with $\mathsf{Q}_*= \{Q_1, Q_2\}$ in Figure \ref{fig:local_1} 
becomes a projective point. In other words, without turning on the twisted masses, the hyperplane 
arrangement of the model remains non-projective and the usage of the Jeffrey-Kirwan 
residue operation cannot be justified. By taking $\eta \in \mathrm{Cone}(\mathsf{Q}_*)$, namely in 
the geometric phase, the projective point $u_*=\lambda$ contributes to the Jeffrey-Kirwan residue in \eqref{lc_yukawa_mon} and we obtain
\begin{align}
Y_{zzz}(z)=\sum_{d=0}^{\infty}z^d
\mathop{\textrm{Res}}_{u=\lambda}\frac{u^3 \left(-u\right)^{2(d-1)}}{(u-\lambda)^{2(d+1)}}
=\frac{1}{1-z}.
\label{res_yukawa1}
\end{align}
In this example the mirror map \eqref{mirror_map} becomes trivial, i.e.
$\log q =\log z$, and the A-model Yukawa coupling \eqref{local_a_yukawa} is given by
\begin{align}
\widetilde{Y}_{hhh}(q)=\frac{1}{1-q}.
\end{align}

\begin{remark}
Turning on the twisted masses in a symmetric manner for the base and fiber directions as
$\lambda_1=\lambda_2=\lambda_3=\lambda_4=\lambda$ to maintain the flop symmetry 
$z \leftrightarrow z^{-1}$ of the resolved conifold interchanging the compact and the 
non-compact directions, we obtain
\begin{align}
Y'_{zzz}(z)=\sum_{d=0}^{\infty}z^d
\mathop{\textrm{Res}}_{u=\lambda}\frac{u^3 \left(-u-\lambda\right)^{2(d-1)}}{(u-\lambda)^{2(d+1)}}
=\frac12+\frac{z}{1-z}.
\end{align}
Interestingly, this expression has a different classical triple intersection number and consistent with the result argued in \cite{Forbes:2005xt}.
\end{remark}

\subsubsection{$\mathcal{O}(0)\oplus\mathcal{O}(-2)\to {\IP}^1$}

Next, we consider the local toric Calabi-Yau threefold $\mathcal{O}(0)\oplus\mathcal{O}(-2)\to {\IP}^1$ described by 
a $U(1)$ GLSM with
\begin{align}
\left(
\begin{array}{cccc}
Q_1&Q_2&Q_3&Q_4\\ \hline
\lambda_1&\lambda_2&\lambda_3&\lambda_4
\end{array}
\right)
=
\left(
\begin{array}{cccc}
1&1&0&-2\\ \hline
\lambda&\lambda&-2\lambda&0
\end{array}
\right).
\label{02P}
\end{align}
Following the criteria in Section \ref{subsec:rule_mass}, a twisted mass parameter $\lambda\ne 0$ has been 
introduced such that $u_*=\lambda$ associated with $\mathsf{Q}_*=\{Q_1, Q_2\}=1$ gives a projective 
point (see Figure \ref{fig:local_1}). 
In particular, according to rule {\bf 2}, we have assigned a twisted 
mass $\lambda_3=-2\lambda$ for the neutral chiral multiplet $\Phi_3$ corresponding to $\mathcal{O}(0)$
in order to satisfy the ``Calabi-Yau condition on the divisor" $\lambda_1+\lambda_2+\lambda_3=0$. 
For $\eta \in \mathrm{Cone}(1)$ in the geometric phase, the projective point $u_*=\lambda$ contributes to 
the Jeffrey-Kirwan residue in \eqref{lc_yukawa_mon} and we obtain
\begin{align}
Y_{zzz}(z)=\sum_{d=0}^{\infty}z^d
\mathop{\textrm{Res}}_{u=\lambda}\frac{u^3 \left(-2u\right)^{2d-1}}{2\lambda(u-\lambda)^{2(d+1)}}
=-\frac{1}{2(1-4z)^2}.
\label{02_coni_yukawa}
\end{align}
This agrees with the Yukawa coupling evaluated in \cite{Forbes:2005xt}.
The mirror map \eqref{mirror_map} becomes
\begin{align}
\log q = \log z +2 \sum_{d=1}^{\infty}\frac{(2d-1)!}{(d!)^2}\,z^d
=\log z-2\log \frac{1+\sqrt{1-4z}}{2},\quad \longleftrightarrow\quad 
z=\frac{q}{(1+q)^2},
\end{align}
and the A-model Yukawa coupling \eqref{local_a_yukawa} is given by
\begin{align}
\widetilde{Y}_{hhh}(q)=-\frac12+\frac{q}{1+q}.
\end{align}

\subsubsection{Local ${\IP}^2$: $\mathcal{O}(-3)\to {\IP}^2$}

As an explicit example with $\dim H_4(X,{\IZ})=1$, we consider the local Calabi-Yau threefold 
$K_{{\IP}^2}$ described by a $U(1)$ GLSM with 
\begin{align}
\left(
\begin{array}{cccc}
Q_1&Q_2&Q_3&Q_4\\ \hline
\lambda_1&\lambda_2&\lambda_3&\lambda_4
\end{array}
\right)
=
\left(
\begin{array}{cccc}
1&1&1&-3\\ \hline
\lambda&\lambda&\lambda&0
\end{array}
\right).
\end{align}
Here we have introduced a twisted mass parameter $\lambda\ne 0$ for the base space ${\IP}^2$ such that 
$u_*=\lambda$ associated with $\mathsf{Q}_*=\{Q_1, Q_2, Q_3\}=1$ provides a projective point (see Figure \ref{fig:local_1}). 
For $\eta \in \mathrm{Cone}(1)$ in the geometric phase, the projective point $u_*=\lambda$ gives a contribution to 
the Jeffrey-Kirwan residue in the GLSM correlator \eqref{lc_yukawa_mon} and we obtain
\begin{align}
Y_{zzz}(z)=\sum_{d=0}^{\infty}z^d
\mathop{\textrm{Res}}_{u=\lambda}\frac{u^3 \left(-3u\right)^{3d-1}}{(u-\lambda)^{3(d+1)}}
=-\frac{1}{3(1+27z)}.
\end{align}
The mirror map \eqref{mirror_map} is given by
\begin{align}
\log q =\log z+3\sum_{d=1}^{\infty}(-z)^d\,\frac{(3d-1)!}{(d!)^3},
\end{align}
and by using (\ref{a_yukawa_3}), the Gromov-Witten invariants computed in \cite{Chiang:1999tz}
can be precisely reproduced.

\subsubsection{Local $F_0$: $\mathcal{O}(-2,-2)\to {\IP}^1\times {\IP}^1$}

\begin{figure}[t]
\centering
\includegraphics[width=140mm]{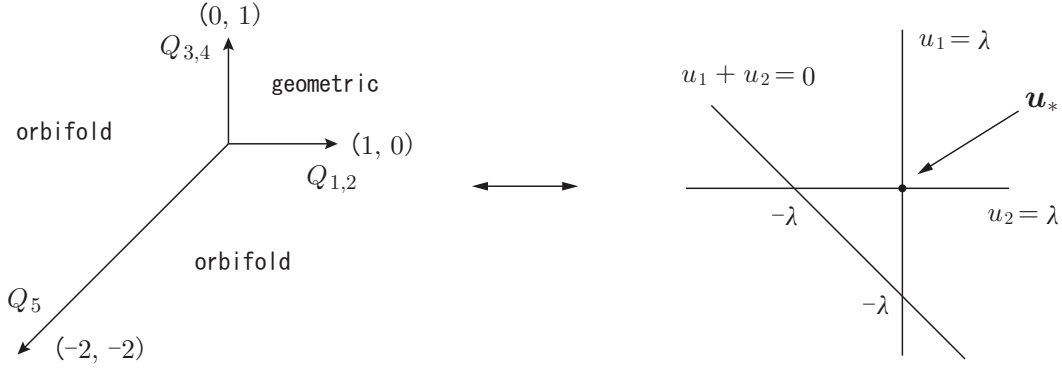}
\caption{The left figure describes the secondary fan of the local $F_0$, and the right figure describes the hyperplanes $\{u_1=\lambda, u_2=\lambda, u_1+u_2=0\}$ corresponding to the charge vectors $\{Q_{1,2}, Q_{3,4}, Q_5\}$. The point $\mathbf{u}_*=(\lambda,\lambda)$ is a projective point associated with $\mathsf{Q}_*=\{Q_{1,2},Q_{3,4}\}$. The orbifold phases in the left figure are described by the orbifold geometry $T^*S^3/{\IZ}_2$.}
\label{fig:local_2}
\end{figure}

As a second example with $\dim H_4(X,{\IZ})=1$, let us consider the local Hirzebruch surface 
$K_{F_0}=K_{{\IP}^1\times {\IP}^1}$ described by a $U(1)^2$ GLSM with
\begin{align}
\left(
\begin{array}{ccccc}
Q_1&Q_2&Q_3&Q_4&Q_5\\ \hline
\lambda_1&\lambda_2&\lambda_3&\lambda_4&\lambda_5
\end{array}
\right)
=
\left(
\begin{array}{ccccc}
1&1&0&0&-2\\
0&0&1&1&-2\\ \hline
\lambda&\lambda&\lambda&\lambda&0
\end{array}
\right).
\end{align}
Here, based on the rules in Section \ref{subsec:rule_mass}, 
we have introduced a twisted mass parameter $\lambda\ne 0$ in a symmetric way for $z_1 \leftrightarrow z_2$ in 
the base space ${\IP}^1\times {\IP}^1$, such that $\mathbf{u}_*=(\lambda,\lambda)$ associated with 
$\mathsf{Q}_*=\{Q_1, Q_2, Q_3, Q_4\}$ provides a projective point (see Figure \ref{fig:local_2}). For 
$\eta \in \mathrm{Cone}(Q_1,Q_1+Q_3)$ inside the geometric phase, the projective point 
$\mathbf{u}_*=(\lambda,\lambda)$ gives a contribution to the Jeffrey-Kirwan residue in the GLSM 
correlation functions in \eqref{lc_yukawa_mon} as
\begin{align}
Y_{z_iz_jz_k}(z_1,z_2)&=\sum_{d_1,d_2=0}^{\infty}z_1^{d_1}z_2^{d_2}
\mathop{\textrm{Res}}_{u_2=\lambda}\mathop{\textrm{Res}}_{u_1=\lambda}
\frac{u_iu_ju_k \left(-2u_1-2u_2\right)^{2d_1+2d_2-1}}
{(u_1-\lambda)^{2(d_1+1)}(u_2-\lambda)^{2(d_2+1)}},
\end{align}
and each component has the following exact form:
\begin{align}
\begin{split}
&
Y_{z_1z_1z_1}(z_1,z_2)=
\frac{(1-4z_2)^2-16z_1(1+z_1)}{4\Delta_{F_0}(z_1,z_2)},\quad
Y_{z_1z_1z_2}(z_1,z_2)=
\frac{16z_1^2-(1-4z_2)^2}{4\Delta_{F_0}(z_1,z_2)},\\
&
Y_{z_1z_2z_2}(z_1,z_2)=
\frac{16z_2^2-(1-4z_1)^2}{4\Delta_{F_0}(z_1,z_2)},\quad
Y_{z_2z_2z_2}(z_1,z_2)=
\frac{(1-4z_1)^2-16z_2(1+z_2)}{4\Delta_{F_0}(z_1,z_2)},
\end{split}
\end{align}
where
\begin{align}
\Delta_{F_0}(z_1,z_2)=1-8(z_1+z_2)+16(z_1-z_2)^2.
\end{align}
The above result is in agreement with the Yukawa couplings evaluated from the local mirror symmetry approach 
in \cite{Forbes:2005xt,Haghighat:2008gw}. The mirror map for $K_{F_0}$ is obtained by \eqref{mirror_map} as
\begin{align}
\log q_i =\log z_i+2\mathop{\sum_{d_1, d_2=0}^{\infty}}\limits_{(d_1,d_2)\ne (0,0)}
\frac{(2d_1+2d_2-1)!}{(d_1!)^2(d_2!)^2}\,z_1^{d_1}z_2^{d_2},\quad
i=1,2,
\end{align}
and then the formula \eqref{a_yukawa_3} correctly reproduces the Gromov-Witten invariants studied in \cite{Chiang:1999tz}.

\subsubsection{Local $F_1$}

\begin{figure}[t]
\centering
\includegraphics[width=140mm]{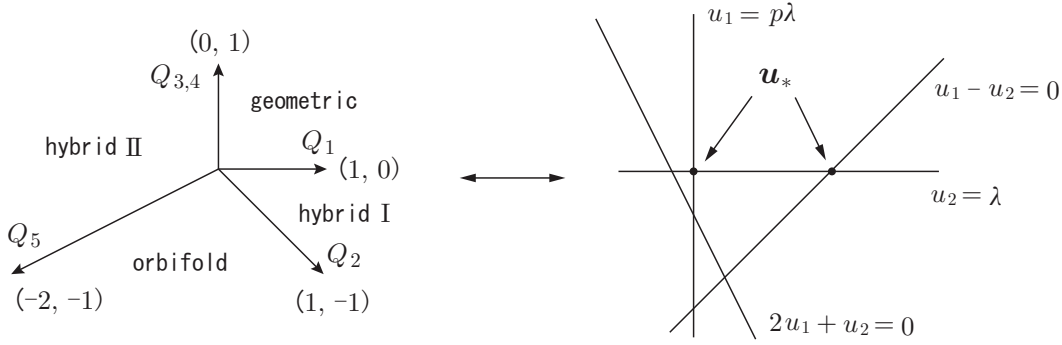}
\caption{The left figure describes the secondary fan of the local $F_1$, and the right figure describes the hyperplanes $\{u_1=p\lambda,u_1-u_2=0, u_2=\lambda, 2u_1+u_2=0\}$ corresponding to the charge vectors $\{Q_{1}, Q_{2}, Q_{3,4}, Q_5\}$. The points $\mathbf{u}_*=(p\lambda,\lambda)$, $(\lambda,\lambda)$ are projective points associated with $\mathsf{Q}_*=\{Q_{1},Q_{3,4}\}$, $\{Q_{2},Q_{3,4}\}$.}
\label{fig:local_3}
\end{figure}

As a third example with $\dim H_4(X,{\IZ})=1$, we consider the local Hirzebruch surface $K_{F_1}$ 
obtained by the one-point blow-up of $K_{{\IP}^2}$, which can be described by a $U(1)^2$ GLSM with
\begin{align}
\left(
\begin{array}{ccccc}
Q_1&Q_2&Q_3&Q_4&Q_5\\ \hline
\lambda_1&\lambda_2&\lambda_3&\lambda_4&\lambda_5
\end{array}
\right)
=
\left(
\begin{array}{ccccc}
1&1&0&0&-2\\
0&-1&1&1&-1\\ \hline
p \lambda&0&\lambda&\lambda&0
\end{array}
\right).
\label{F1_charge}
\end{align}
Here we have introduced a twisted mass parameter $\lambda\ne 0$ and a free parameter $p\neq -1/2$ for the base space $F_1$, 
such that $\mathbf{u}_*=(p \lambda, \lambda)$ and $(\lambda, \lambda)$ associated with 
$\mathsf{Q}_*=\{Q_{1}, Q_3, Q_4\}$ and $\{Q_{2}, Q_3, Q_4\}$ provide projective points respectively (see Figure \ref{fig:local_3}). Note that 
if we choose $p=-1/2$, the point $\mathbf{u}_*=(p \lambda, \lambda)$ becomes non-projective. 
For $\eta \in \mathrm{Cone}(Q_1,Q_1+Q_3)$ in the geometric phase, the projective points 
$\mathbf{u}_*=(p \lambda, \lambda)$, $(\lambda, \lambda)$ contribute to the Jeffrey-Kirwan residue in \eqref{lc_yukawa_mon} as
\begin{align}
Y_{z_iz_jz_k}(z_1,z_2)&=\sum_{d_1,d_2=0}^{\infty}z_1^{d_1}z_2^{d_2}
\left(\mathop{\textrm{Res}}_{u_2=\lambda}\mathop{\textrm{Res}}_{u_1=p\lambda}
+\mathop{\textrm{Res}}_{u_1=\lambda}\mathop{\textrm{Res}}_{u_2=\lambda}\right)
\nonumber\\
&\qquad\qquad \times
\frac{u_iu_ju_k \left(-2u_1-u_2\right)^{2d_1+d_2-1}}
{(u_1-p\lambda)^{d_1+1}(u_2-\lambda)^{2(d_2+1)}(u_1-u_2)^{d_1-d_2+1}},
\end{align}
and we obtain
\begin{align}
\begin{split}
&
Y_{z_1z_1z_1}(z_1,z_2)=-6x+\frac13
+\frac{-1-4z_1^2+z_2-z_1(7-6z_2)}{3\Delta_{F_1}(z_1,z_2)},\\
&
Y_{z_1z_1z_2}(z_1,z_2)=12x-\frac23
+\frac{-1+8z_1^2+z_2+z_1(2-3z_2)}{3\Delta_{F_1}(z_1,z_2)},\\
&
Y_{z_1z_2z_2}(z_1,z_2)=-24x+\frac43
+\frac{z_2(1-12z_1)-(1-4z_1)^2}{3\Delta_{F_1}(z_1,z_2)},\\
&
Y_{z_2z_2z_2}(z_1,z_2)=48x-\frac83
+\frac{2(1-4z_1)^2+z_2(1+60z_1)}{3\Delta_{F_1}(z_1,z_2)},
\label{f1_yukawa}
\end{split}
\end{align}
where
\begin{align}
\Delta_{F_1}(z_1,z_2)=(1-4z_1)^2-z_2(1-36z_1+27z_1z_2),
\end{align}
and 
\begin{align}
x = \frac{(p+1)(3p+1)}{18(2p+1)^2}.
\label{fj_local_F1}
\end{align}
The above expressions completely agree with the result in \cite{Forbes:2005xt} (and \cite{Haghighat:2008gw} for 
$p=0$ or equivalently $x=1/18$) and we also see that the free parameter $x$ arising from a degree of freedom of 
the twisted mass of a chiral matter multiplet reproduces the ``moduli parameter'' discussed in \cite{Forbes:2005xt}.

\begin{remark}
In addition to the twisted masses represented in \eqref{F1_charge}, we can also turn on a non-zero value for 
$\lambda_2$ as $\lambda_1=p \lambda$, $\lambda_2=q \lambda$, $\lambda_3=\lambda_4=\lambda$, $\lambda_5=0$ 
to make the poles associated with $\mathsf{Q}_*=\{Q_{1}, Q_3, Q_4\}$ and 
$\{Q_{2}, Q_3, Q_4\}$ projective. 
Then we obtain the same result with \eqref{f1_yukawa} while $x$ is defined by
\begin{align}
x = \frac{(2q+3)^2p^2+(4q^2+15q+12)p+(q+1)(q+3)}{6(2p+1)^2(2q+3)^2}.
\end{align}
This means that additional mass deformation is possible in this case, but does not introduce an independent 
moduli parameter.
\end{remark}

From \eqref{mirror_map}, one can easily show that the mirror map for $K_{F_1}$ is given by
\begin{align}
\log q_i =\log z_i+c_i \mathop{\sum_{d_1, d_2=0}^{\infty}}\limits_{(d_1,d_2)\ne (0,0)}
\frac{(-1)^{d_2}(2d_1+3d_2-1)!}{d_1!(d_2!)^2(d_1+d_2)!}\,
z_1^{d_1+d_2} z_2^{d_2},\quad
i=1,2,
\end{align}
where $c_1=2$ and $c_2=1$. Combining this with the formula in \eqref{a_yukawa_3}, the Gromov-Witten invariants studied in \cite{Chiang:1999tz} can be appropriately reproduced.

\subsubsection{Local $F_2$}

\begin{figure}[t]
\centering
\includegraphics[width=140mm]{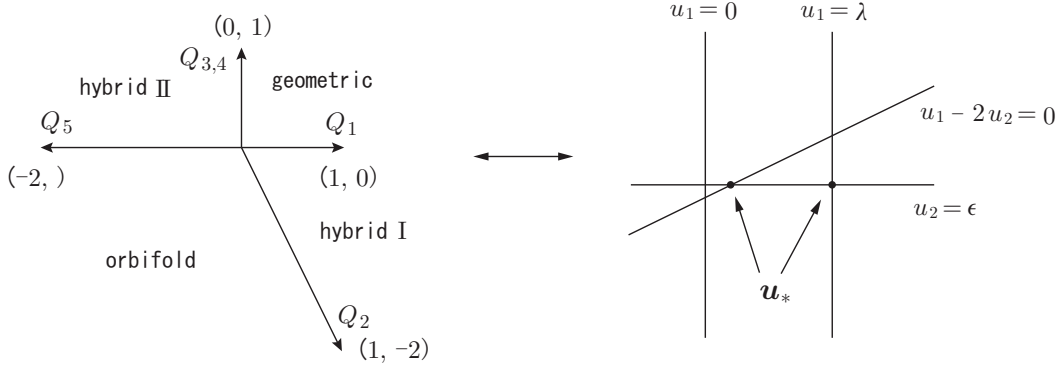}
\caption{The left figure describes the secondary fan of the local $F_2$, and the right figure describes the hyperplanes $\{u_1=\lambda, u_1-2u_2=0, u_2=\epsilon, u_1=0\}$ corresponding to the charge vectors $\{Q_{1}, Q_2,  Q_{3,4}, Q_5\}$. The points $\mathbf{u}_*=(\lambda,\epsilon)$, $(2\epsilon,\epsilon)$ are projective points associated with $\mathsf{Q}_*=\{Q_{1},Q_{3,4}\}$, $\{Q_{2},Q_{3,4}\}$. The orbifold phase in the left figure is described by the orbifold geometry ${\IC}^3/{\IZ}_4$.}
\label{fig:local_4}
\end{figure}

As a fourth example with $\dim H_4(X,{\IZ})=1$, let us study the local Hirzebruch surface $K_{F_2}$ 
(a local $A_1$ geometry) described by a $U(1)^2$ GLSM with
\begin{align}
\left(
\begin{array}{ccccc}
Q_1&Q_2&Q_3&Q_4&Q_5\\ \hline
\lambda_1&\lambda_2&\lambda_3&\lambda_4&\lambda_5
\end{array}
\right)
=
\left(
\begin{array}{ccccc}
1&1&0&0&-2\\
0&-2&1&1&0\\ \hline
\lambda&0&\epsilon&\epsilon&0
\end{array}
\right).
\label{F2_charge}
\end{align}
Here we have introduced twisted mass parameters $\epsilon, \lambda\ne 0$ with $\epsilon\ne \lambda$ for compact 
directions such that $\mathbf{u}_*=(\lambda,\epsilon)$ and $(2\epsilon,\epsilon)$ associated with 
$\mathsf{Q}_*=\{Q_{1}, Q_3, Q_4\}$ and $\{Q_{2}, Q_3, Q_4\}$ provide the projective points respectively (see Figure \ref{fig:local_4}).

In this model, the chiral multiplet $\Phi_5$, which is neutral under the second $U(1)$ gauge symmetry describes 
the non-compact fiber coordinate $X_5$, and 
the compact divisor $\{X_5=0\}$ (i.e. $F_2$) contains the blow-up mode of the ${\IZ}_2$ singularity.
Therefore, following rule {\bf 2} in Section \ref{subsec:rule_mass}, we impose the ``Calabi-Yau condition on the 
divisor'' as $\lambda_2+\lambda_3+\lambda_4+\lambda_5=0$, which implies 
the need to take a limit $\epsilon \to 0$ in the final
step.
For $\eta \in \mathrm{Cone}(Q_1,Q_1+Q_3)$ in the geometric phase, the projective points 
$\mathbf{u}_*=(\lambda,\epsilon)$ and $(2\epsilon,\epsilon)$ contribute to the Jeffrey-Kirwan residue in \eqref{lc_yukawa_mon} as
\begin{align}
Y_{z_iz_jz_k}(z_1,z_2)&=\lim_{\epsilon \to 0}
\sum_{d_1,d_2=0}^{\infty}z_1^{d_1}z_2^{d_2}
\left(\mathop{\textrm{Res}}_{u_2=\epsilon}\mathop{\textrm{Res}}_{u_1=\lambda}
+\mathop{\textrm{Res}}_{u_1=2\epsilon}\mathop{\textrm{Res}}_{u_2=\epsilon}\right)
\nonumber\\
&\qquad\qquad\qquad \times
\frac{u_iu_ju_k \left(-2u_1\right)^{2d_1-1}}
{(u_1-\lambda)^{d_1+1}(u_2-\epsilon)^{2(d_2+1)}(u_1-2u_2)^{d_1-2d_2+1}},
\end{align}
and we obtain
\begin{align}
\begin{split}
&
Y_{z_1z_1z_1}(z_1,z_2)=
\frac{-1}{\Delta_{F_2}(z_1,z_2)},\quad
Y_{z_1z_1z_2}(z_1,z_2)=
\frac{2z_1-\frac12}{\Delta_{F_2}(z_1,z_2)},\\
&
Y_{z_1z_2z_2}(z_1,z_2)=
\frac{z_2(1-8z_1)}{(1-4z_2)\Delta_{F_2}(z_1,z_2)},\quad
Y_{z_2z_2z_2}(z_1,z_2)=
\frac{z_2(24z_1z_2+2z_1-2z_2-\frac12)}{(1-4z_2)^2\Delta_{F_2}(z_1,z_2)},
\end{split}
\end{align}
where
\begin{align}
\Delta_{F_2}(z_1,z_2)=(1-4z_1)^2-64z_1^2z_2.
\end{align}
The above expressions indeed agree with the result obtained in \cite{Forbes:2005xt}.
The mirror map \eqref{mirror_map} for $K_{F_2}$ is given by
\begin{align}
\log q_i =\log z_i+c_i \log \frac{1+\sqrt{1-4z_2}}{2}
+2\delta_{i,1}\mathop{\sum_{d_1, d_2=0}^{\infty}}\limits_{(d_1,d_2)\ne (0,0)}
\frac{(2d_1+4d_2-1)!}{d_1!(d_2!)^2(d_1+2d_2)!}\,
z_1^{d_1+2d_2}z_2^{d_2},
\end{align}
where $i=1,2$, and $(c_1, c_2) =(1, -2)$.  The Gromov-Witten invariants computed in \cite{Chiang:1999tz} 
from the local mirror symmetry approach can also be reproduced by using \eqref{a_yukawa_3}.

\subsubsection{Local $dP_2$}

As a last example with $\dim H_4(X,{\IZ})=1$, 
we consider the local del Pezzo surface $K_{dP_2}$, which can be obtained by a one-point blow-up of $K_{F_0}$ 
or $K_{F_1}$. This local toric variety can be described by a $U(1)^3$ GLSM with
\begin{align}
\left(
\begin{array}{cccccc}
Q_1&Q_2&Q_3&Q_4&Q_5&Q_6\\ \hline
\lambda_1&\lambda_2&\lambda_3&\lambda_4&\lambda_5&\lambda_6
\end{array}
\right)
=
\left(
\begin{array}{cccccc}
1&-1&1&0&0&-1\\
-1&1&0&0&1&-1\\ 
0&1&-1&1&0&-1\\ \hline
\lambda&q\lambda&\lambda&p\lambda&p\lambda&0
\end{array}
\right).
\end{align}
Here we have introduced a twisted mass parameter $\lambda\ne 0$ and two free parameters $p$ and $q$ 
respecting the interchanging symmetry $z_2 \leftrightarrow z_3$ for the base space $dP_2$,
such that $\mathbf{u}_*=((p+1)\lambda,p \lambda,p \lambda)$, 
$((p+1)\lambda,p \lambda,(q+1)\lambda)$, $((p+1)\lambda,(q+1)\lambda,p \lambda)$, and $((q+2)\lambda,(q+1)\lambda,(q+1)\lambda)$ associated with 
$\mathsf{Q}_*=\{Q_1, Q_3, Q_4, Q_5\}$, 
$\{Q_1, Q_2, Q_5\}$, $\{Q_2, Q_3, Q_4\}$, and $\{Q_1, Q_2, Q_3\}$ provide the projective points, respectively. 
Note that the particular values for $p$ and $q$ satisfying $3p+1=0$, $2p+q+2=0$, $3q+4=0$ should be excluded 
in order to maintain the projective condition. 

For $\eta=(5,3,2)$ in the geometric phase, we find that the above projective points contribute to the Jeffrey-Kirwan residue 
in \eqref{lc_yukawa_mon} as
\begin{align}
Y_{z_iz_jz_k}(z_1,z_2,z_3)&=
\sum_{d_1,d_2,d_3=0}^{\infty}z_1^{d_1}z_2^{d_2}z_3^{d_3}
\Big(\mathop{\textrm{Res}}_{u_1=(p+1)\lambda}
\mathop{\textrm{Res}}_{u_3=p \lambda}\mathop{\textrm{Res}}_{u_2=p \lambda}
+\mathop{\textrm{Res}}_{u_3=(q+1)\lambda}\mathop{\textrm{Res}}_{u_1=(p+1)\lambda}
\mathop{\textrm{Res}}_{u_2=p \lambda}
\nonumber\\
&\qquad\qquad
+\mathop{\textrm{Res}}_{u_2=(q+1)\lambda}\mathop{\textrm{Res}}_{u_3= p \lambda}
\mathop{\textrm{Res}}_{u_1=u_3+\lambda}
+\mathop{\textrm{Res}}_{u_3=(q+1)\lambda}\mathop{\textrm{Res}}_{u_1=u_3+\lambda}
\mathop{\textrm{Res}}_{u_2=u_1-u_3+q\lambda}\Big)
\nonumber\\
&\ \ \ \times
\frac{u_iu_ju_k \left(-u_1-u_2-u_3\right)^{d_1+d_2+d_3-1}}
{(u_1-u_2-\lambda)^{d_1-d_2+1}(-u_1+u_2+u_3-q\lambda)^{-d_1+d_2+d_3+1}}
\nonumber\\
&\ \ \ \times
\frac{1}{(u_1-u_3-\lambda)^{d_1-d_3+1}
(u_3-p \lambda)^{d_3+1}(u_2-p \lambda)^{d_2+1}}.
\label{loc_dp2_correl}
\end{align}
Note that the number of intersecting hyperplanes at the point $\mathbf{u}_*=((p+1)\lambda,p \lambda,p \lambda)$ is larger
than $r=3$. To deal with this ``degenerate point", which requires careful treatment of the order of the iterated residue, 
we have applied Theorem \ref{thm:jk_iterate} for a flag $F$ with $\kappa^F=(Q_5, Q_4+Q_5, Q_1+Q_3+Q_4+Q_5)$ and $\nu(F)=1$
as the only constituent in $\mathcal{F}\mathcal{L}^{+}(\mathsf{Q}_*, \eta)$ for 
$\mathsf{Q}_*=\{Q_1, Q_3, Q_4, Q_5\}$.

As a result, we finally obtain
\begin{align}
\begin{split}
&
Y_{z_1z_1z_1}(z_1,z_2,z_3)=6x+2y-1+Y_{z_1z_1z_1}(z_1,z_2,z_3)\big|_{p=q=0},\\
&
Y_{z_1z_1z_2}(z_1,z_2,z_3)=-3x-y+\frac12+Y_{z_1z_1z_2}(z_1,z_2,z_3)\big|_{p=q=0},\\
&
Y_{z_1z_2z_2}(z_1,z_2,z_3)=x+y+Y_{z_1z_2z_2}(z_1,z_2,z_3)\big|_{p=q=0},\\
&
Y_{z_1z_2z_3}(z_1,z_2,z_3)=2x-\frac12+Y_{z_1z_2z_3}(z_1,z_2,z_3)\big|_{p=q=0},\\
&
Y_{z_2z_2z_2}(z_1,z_2,z_3)=-y-\frac14+Y_{z_2z_2z_2}(z_1,z_2,z_3)\big|_{p=q=0},\\
&
Y_{z_2z_2z_3}(z_1,z_2,z_3)=-x+\frac14+Y_{z_2z_2z_3}(z_1,z_2,z_3)\big|_{p=q=0},
\end{split}
\end{align}
and
\begin{align}
\begin{split}
&
Y_{z_1z_1z_3}(z_1,z_2,z_3)=Y_{z_1z_1z_2}(z_1,z_3,z_2),\quad Y_{z_1z_3z_3}(z_1,z_2,z_3)=Y_{z_1z_2z_2}(z_1,z_3,z_2),\\
&
Y_{z_2z_2z_2}(z_1,z_2,z_3)=Y_{z_3z_3z_3}(z_1,z_3,z_2),\quad Y_{z_2z_2z_3}(z_1,z_2,z_3)=Y_{z_2z_3z_3}(z_1,z_3,z_2),
\end{split}
\end{align}
with $z_2 \leftrightarrow z_3$ symmetry. Here
\begin{align}
\begin{split}
x&=\frac{\left(12 p^3+33 p^2+19 p+3\right) q+2 \left(8 p^3+14 p^2+7 p+1\right)+(3 p q+q)^2}{(3 p+1)^2 (3 q+4) (2 p+q+2)},
\\
y&=-\frac{5 p^3 (3 q+4)+2 p^2 (3 q+4)^2+2 p \left(6 q^2+13 q+7\right)+2 (q+1)^2}{(3 p+1)^2 (3 q+4) (2 p+q+2)},
\end{split}
\end{align}
and the analytic expressions of $Y_{z_iz_jz_k}(z_1,z_2,z_3)\big|_{p=q=0}$ are summarized in Appendix \ref{app:dp2}. 
Note that $p=q=0$ corresponds to $(x,y)=(1/4, -1/4)$.

The above results completely agree with the Yukawa couplings with the ``moduli parameters'' $x$ and $y$ 
given in \cite{Forbes:2005xt}.
The mirror map \eqref{mirror_map} for $K_{dP_2}$ is given by
\begin{align}
\log q_i =\log z_i+
\mathop{\sum_{d_1,d_2,d_3=0}^{\infty}}\limits_{(d_1,d_2,d_3)\ne (0,0,0)}
\frac{(-1)^{d_1}(3d_1+2d_2+2d_3-1)!}{d_1!d_2!d_3!(d_1+d_2)!(d_1+d_3)!}\,
z_1^{d_1+d_2+d_3}z_2^{d_1+d_2}z_3^{d_1+d_3},
\end{align}
where $i=1,2,3$, and the associated Gromov-Witten invariants studied in \cite{Chiang:1999tz} are correctly 
reproduced by using \eqref{a_yukawa_3}.

\subsubsection{Local $A_2$ geometry}

\begin{figure}[t]
\centering
\includegraphics[width=120mm]{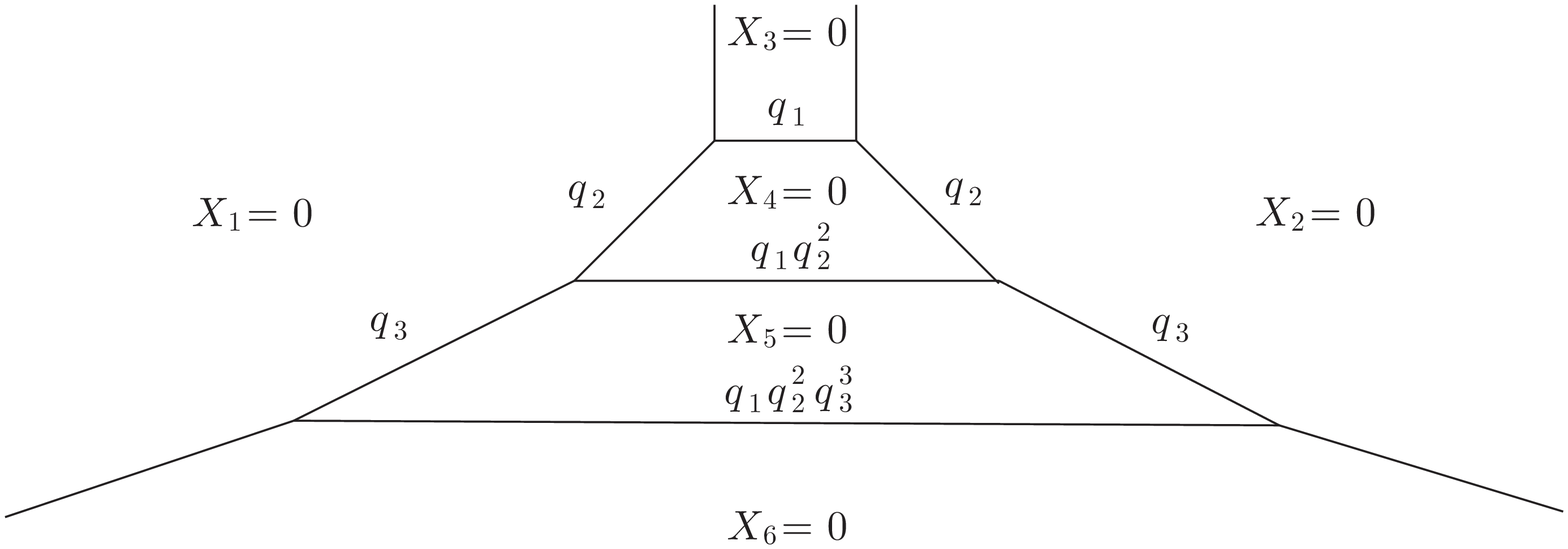}
\caption{The toric web diagram for the local $A_2$ geometry. $X_i$, $i=1,\ldots,6$ are homogeneous local 
coordinates corresponding to the chiral multiplets $\Phi_i$, and $\{X_4=0\}$ or $\{X_5=0\}$ gives a compact divisor.
The parameters $q_1$, $q_2$ and $q_3$ are exponentiated K\"ahler moduli associated with the $U(1)^3$ gauge group.}
\label{fig:local_A2}
\end{figure}

Let us consider a local $A_2$ geometry with $\dim H_4(X,{\IZ})=2$, which is a fibered $A_2$ geometry over ${\IP}^1$ 
represented in Figure \ref{fig:local_A2}. This geometry engineers the four dimensional pure $SU(3)$ gauge theory and
can be described by a $U(1)^3$ GLSM with \cite{Katz:1996fh,Katz:1997eq,Chiang:1999tz}
\begin{align}
\left(
\begin{array}{cccccc}
Q_1&Q_2&Q_3&Q_4&Q_5&Q_6\\ \hline
\lambda_1&\lambda_2&\lambda_3&\lambda_4&\lambda_5&\lambda_6
\end{array}
\right)
=
\left(
\begin{array}{cccccc}
1&1&-2&0&0&0\\
0&0&1&-2&1&0\\ 
0&0&0&1&-2&1\\ \hline
\epsilon&\epsilon&0&0&0&\lambda
\end{array}
\right).
\end{align}
Here we have introduced the twisted mass parameters $\epsilon, \lambda\ne 0$ with $\epsilon\ne \lambda$ for a projective hyperplane arrangement 
in the geometric phase with $\mathsf{Q}_*=\{Q_1, Q_2, Q_3, Q_6\}$, 
$\{Q_1, Q_2, Q_5, Q_6\}$ and $\{Q_1, Q_2, Q_3, Q_4\}$. 

Following rule {\bf 1} in Section \ref{subsec:rule_mass}, we have not included the mass parameters for the 
chiral multiplets $\Phi_3$, $\Phi_4$, and $\Phi_5$ with charge vectors $Q_3$, $Q_4$, and $Q_5$ which blow up 
the singularities.
Furthermore, the chiral multiplet $\Phi_4$, which is neutral with respect to the first $U(1)$ gauge symmetry, describes 
a non-compact fiber coordinate $X_4$ and the compact divisor $\{X_4=0\}$ contains the blow-up mode of the 
${\IZ}_2$ singularity. 
Therefore, by following rule {\bf 2}, we impose the ``Calabi-Yau condition on the divisor'' as
$\lambda_1+\lambda_2+\lambda_3+\lambda_4=0$, which implies 
the need to take a limit $\epsilon \to 0$ in the final step.

By taking e.g. $\eta=(2,1,2)$ in the geometric phase, the projective points 
$\mathbf{u}_*=(\epsilon, 2\epsilon, \lambda)$, $(\epsilon, 2\lambda, \lambda)$, and $(\epsilon, 2\epsilon, 4\epsilon)$ 
contribute to the Jeffrey-Kirwan residue in the GLSM correlators \eqref{lc_yukawa_mon} as
\begin{align}
Y_{z_iz_jz_k}(z_1,z_2,z_3)&=\lim_{\epsilon \to 0}
\sum_{d_1,d_2,d_3=0}^{\infty}z_1^{d_1}z_2^{d_2}z_3^{d_3}
\left(\mathop{\textrm{Res}}_{u_2=2\epsilon}
\mathop{\textrm{Res}}_{u_3=\lambda}\mathop{\textrm{Res}}_{u_1=\epsilon}
+\mathop{\textrm{Res}}_{u_2=2\lambda}\mathop{\textrm{Res}}_{u_1=\epsilon}
\mathop{\textrm{Res}}_{u_3=\lambda}
+\mathop{\textrm{Res}}_{u_3=4\epsilon}\mathop{\textrm{Res}}_{u_2=2\epsilon}
\mathop{\textrm{Res}}_{u_1=\epsilon}\right)
\nonumber\\
&\qquad\qquad\qquad \times
\frac{u_iu_ju_k \left(-2u_2+u_3\right)^{2d_2-d_3-1}
\left(u_2-2u_3\right)^{-d_2+2d_3-1}}
{(u_1-\epsilon)^{2(d_1+1)}(-2u_1+u_2)^{-2d_1+d_2+1}(u_3-\lambda)^{d_3+1}},
\label{loc_A2_correl}
\end{align}
and we obtain the exact expressions represented in Appendix \ref{app:a2}.
The mirror map \eqref{mirror_map} is given by
\begin{align}
\begin{split}
\log q_1&= \log z_1 - 2\log \frac{1+\sqrt{1-4z_1}}{2},
\quad \longleftrightarrow\quad 
z_1=\frac{q_1}{(1+q_1)^2},\\
\log q_2&= \log z_2 + \log \frac{1+\sqrt{1-4z_1}}{2}+2M_1(z_1,z_2,z_3)
-M_2(z_1,z_2,z_3),\\
\log q_3&= \log z_3 - M_1(z_1,z_2,z_3)
+2M_2(z_1,z_2,z_3),
\end{split}
\end{align}
where
\begin{align}
\begin{split}
M_1(z_1,z_2,z_3)&=
\mathop{\sum_{d_1,d_2,d_3=0}^{\infty}}\limits_{(d_1,d_2,d_3)\ne (0,0,0)}
\frac{(-1)^{d_3}(2d_2+3d_3-1)!}{(d_1!)^2d_2!d_3!(-2d_1+d_2+2d_3)!}\,
z_1^{d_1}z_2^{d_2+2d_3}z_3^{d_3},\\
M_2(z_1,z_2,z_3)&=
\mathop{\sum_{d_1,d_2,d_3=0}^{\infty}}\limits_{(d_1,d_2,d_3)\ne (0,0,0)}
\frac{(-1)^{d_2}(6d_1+3d_2+2d_3-1)!}{(d_1!)^2d_2!d_3!(4d_1+2d_2+d_3)!}\,
z_1^{d_1}z_2^{2d_1+d_2}z_3^{4d_1+2d_2+d_3}.
\end{split}
\end{align}
From \eqref{a_yukawa_3} we can reproduce the Gromov-Witten invariants studied in Table 4 of \cite{Chiang:1999tz}, 
except a one coefficient $n_{1,0,0}=-2/3$. A similar fractional number also appears at the degree 
$(0,1)$ invariant $n_{0,1}=-1/2$ for the local $F_2$ that 
we have investigated in this section (see also Table 11 in \cite{Chiang:1999tz}).

\subsection{Local toric Calabi-Yau fourfolds}\label{subsec:local_cy4}

Finally we focus on the local nef toric Calabi-Yau fourfolds whose exact properties have been 
studied in \cite{Klemm:2007in}. We again confirm that the localization formula for the A-twisted 
GLSM correlation functions provides the exact expressions for local B-model Yukawa couplings 
appropriately, with the aid of our formalism.

\subsubsection{$\mathcal{O}(-1)\oplus \mathcal{O}(-2)\to {\IP}^2$}

\begin{figure}[t]
\centering
\includegraphics[width=150mm]{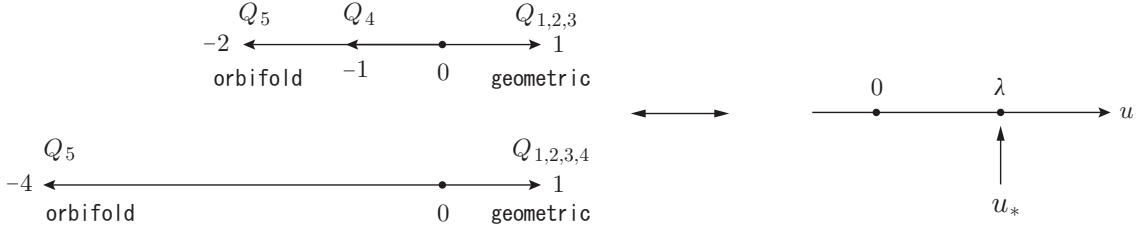}
\caption{The first (resp. second) figure in the left describes the secondary fan of $\mathcal{O}(-1)\oplus \mathcal{O}(-2)\to {\IP}^2$ (resp. $\mathcal{O}(-4)\to {\IP}^3$), and the right figure describes the hyperplanes $\{u=\lambda, u=0\}$ corresponding to the charge vectors $\{Q_{1,2,3}, Q_{4,5}\}$ (resp. $\{Q_{1,2,3,4}, Q_{5}\}$). The point $u_*=\lambda$ is a projective point associated with $\mathsf{Q}_*=Q_{1,2,3}=1$ (resp. $\mathsf{Q}_*=Q_{1,2,3,4}=1$).}
\label{fig:local_5}
\end{figure}

Let us consider the local toric Calabi-Yau fourfold $\mathcal{O}(-1)\oplus \mathcal{O}(-2)\to {\IP}^2$ 
described by a $U(1)$ GLSM with
\begin{align}
\left(
\begin{array}{ccccc}
Q_1&Q_2&Q_3&Q_4&Q_5\\ \hline
\lambda_1&\lambda_2&\lambda_3&\lambda_4&\lambda_5
\end{array}
\right)
=
\left(
\begin{array}{ccccc}
1&1&1&-1&-2\\ \hline
\lambda&\lambda&\lambda&0&0
\end{array}
\right),
\end{align}
where, following the rules in Section \ref{subsec:rule_mass}, 
we have introduced a twisted mass parameter $\lambda\ne 0$ for the base space ${\IP}^2$ such 
that $u_*=\lambda$ associated with $\mathsf{Q}_*=\{Q_1, Q_2, Q_3\}=1$ gives a projective point 
(see Figure \ref{fig:local_5}). For $\eta \in \mathrm{Cone}(1)$ in the geometric phase, the projective 
point $u_*=\lambda$ contributes to the Jeffrey-Kirwan residue in the GLSM correlation function 
\eqref{lc_yukawa_mon} as
\begin{align}
Y_{zzzz}(z)=\sum_{d=0}^{\infty}z^d
\mathop{\textrm{Res}}_{u=\lambda}\frac{u^4 \left(-u\right)^{d-1}\left(-2u\right)^{2d-1}}
{(u-\lambda)^{3(d+1)}}
=\frac{1}{2(1+4z)}.
\end{align}
In this example, the mirror map \eqref{mirror_map} is trivial, i.e. $\log q=\log z$, and the A-model 
Yukawa coupling \eqref{local_a_yukawa} takes a form
\begin{align}
\widetilde{Y}_{hhhh}(q)=\frac{1}{2(1+4q)}.
\end{align}
One finds that the identity
\begin{align}
\frac{1}{1-4q}=
\left(\sum_{d=0}^{\infty}\binom{2d}{d}\,q^d\right)^2,
\end{align}
is equivariant to 
the factorization \eqref{factorize:yukawa_4} of the four-point Yukawa coupling $\widetilde{Y}_{hhhh}(q)$ into 
the three-point Yukawa coupling $\widetilde{Y}_{hhh^2}(q)$ in \cite{Klemm:2007in}:
\begin{align}
\widetilde{Y}_{hhhh}(q)=2\widetilde{Y}_{hhh^2}(q)^2, \qquad 
\widetilde{Y}_{hhh^2}(q)=\sum_{d=0}^{\infty}\frac{1}{2}\binom{2d}{d}\,(-q)^d.
\end{align}

\subsubsection{Local ${\IP}^3$: $\mathcal{O}(-4)\to {\IP}^3$}

Next, let us consider the local toric Calabi-Yau fourfold $K_{{\IP}^3}$ described by a 
$U(1)$ GLSM with
\begin{align}
\left(
\begin{array}{ccccc}
Q_1&Q_2&Q_3&Q_4&Q_5\\ \hline
\lambda_1&\lambda_2&\lambda_3&\lambda_4&\lambda_5
\end{array}
\right)
=
\left(
\begin{array}{ccccc}
1&1&1&1&-4\\ \hline
\lambda&\lambda&\lambda&\lambda&0
\end{array}
\right),
\end{align}
where a twisted mass parameter $\lambda\ne 0$ has been introduced such that $u_*=\lambda$ associated 
with $\mathsf{Q}_*=\{Q_1, Q_2, Q_3, Q_4\}=1$ provides a projective point (see Figure \ref{fig:local_5}). 
For $\eta \in \mathrm{Cone}(1)$ in the geometric phase, the projective point $u_*=\lambda$ contributes to 
the Jeffrey-Kirwan residue in \eqref{lc_yukawa_mon} and we obtain
\begin{align}
Y_{zzzz}(z)=\sum_{d=0}^{\infty}z^d
\mathop{\textrm{Res}}_{u=\lambda}\frac{u^4 \left(-4u\right)^{4d-1}}
{(u-\lambda)^{4(d+1)}}
=-\frac{1}{4\left(1-4^4z\right)}.
\end{align}
The mirror map \eqref{mirror_map} for $K_{{\IP}^3}$ is given by
\begin{align}
\log q =\log z+4\sum_{d=1}^{\infty}\frac{(4d-1)!}{(d!)^4}\,z^d,
\end{align}
and one can check that the factorization \eqref{factorize:yukawa_4} into the three-point Yukawa couplings correctly reproduces the 
previous results for the Gromov-Witten invariants in \cite{Klemm:2007in}.

\subsubsection{$\mathcal{O}(-1,-1)\oplus
\mathcal{O}(-1,-1)\to {\IP}^1\times {\IP}^1$}

\begin{figure}[t]
\centering
\includegraphics[width=120mm]{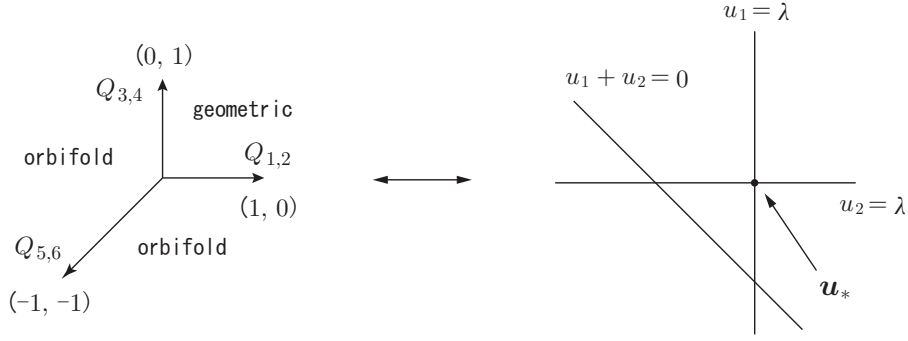}
\caption{The left figure describes the secondary fan of $\mathcal{O}(-1,-1)\oplus
\mathcal{O}(-1,-1)\to {\IP}^1\times {\IP}^1$, and the right figure describes the hyperplanes $\{u_1=\lambda, u_2=\lambda, u_1+u_2=0\}$ corresponding to the charge vectors $\{Q_{1,2}, Q_{3,4}, Q_{5,6}\}$. The point $\mathbf{u}_*=(\lambda,\lambda)$ is a projective point associated with $\mathsf{Q}_*=\{Q_{1,2},Q_{3,4}\}$.}
\label{fig:local_6}
\end{figure}

Finally, we consider the local toric Calabi-Yau fourfold $\mathcal{O}(-1,-1)\oplus
\mathcal{O}(-1,-1)\to {\IP}^1\times {\IP}^1$ described by a $U(1)^2$ GLSM with
\begin{align}
\left(
\begin{array}{cccccc}
Q_1&Q_2&Q_3&Q_4&Q_5&Q_6\\ \hline
\lambda_1&\lambda_2&\lambda_3&\lambda_4&\lambda_5&\lambda_6
\end{array}
\right)
=
\left(
\begin{array}{cccccc}
1&1&0&0&-1&-1\\
0&0&1&1&-1&-1\\ \hline
\lambda&\lambda&\lambda&\lambda&0&0
\end{array}
\right),
\end{align}
where we have introduced a twisted mass parameter $\lambda\ne 0$ in 
a symmetric way for $z_1 \leftrightarrow z_2$ 
in the base space ${\IP}^1\times {\IP}^1$, such that $\mathbf{u}_*=(\lambda,\lambda)$ associated with 
$\mathsf{Q}_*=\{Q_1, Q_2, Q_3, Q_4\}$ gives a projective point (see Figure \ref{fig:local_6}). 
For $\eta \in \mathrm{Cone}(Q_1,Q_1+Q_3)$ in the geometric phase, the projective point 
$\mathbf{u}_*=(\lambda,\lambda)$ gives a contribution to the Jeffrey-Kirwan residue in the GLSM correlators \eqref{lc_yukawa_mon} as
\begin{align}
Y_{z_iz_jz_kz_l}(z_1,z_2)&=\sum_{d_1,d_2=0}^{\infty}z_1^{d_1}z_2^{d_2}
\mathop{\textrm{Res}}_{u_2=\lambda}\mathop{\textrm{Res}}_{u_1=\lambda}
\frac{u_iu_ju_ku_l \left(-u_1-u_2\right)^{2(d_1+d_2-1)}}
{(u_1-\lambda)^{2(d_1+1)}(u_2-\lambda)^{2(d_2+1)}}.
\end{align}
As a result, we obtain
\begin{align}
\begin{split}
&
Y_{z_1z_1z_1z_1}(z_1,z_2)=
\frac{-5(1-z_2)^2+z_1(10+11z_1+10z_2)}{8\Delta_{F_0}'(z_1,z_2)},\\
&
Y_{z_1z_1z_1z_2}(z_1,z_2)=
\frac{(1-z_2)^2+z_1(6-7z_1-10z_2)}{8\Delta_{F_0}'(z_1,z_2)},\\
&
Y_{z_1z_1z_2z_2}(z_1,z_2)=
\frac{3-6(z_1+z_2)+3z_1^2+10z_1z_2+3z_2^2}{8\Delta_{F_0}'(z_1,z_2)},\\
&
Y_{z_1z_2z_2z_2}(z_1,z_2)=
\frac{(1-z_1)^2+z_2(6-10z_1-7z_2)}{8\Delta_{F_0}'(z_1,z_2)},\\
&
Y_{z_2z_2z_2z_2}(z_1,z_2)=
\frac{-5(1-z_1)^2+z_2(10+10z_1+11z_2)}{8\Delta_{F_0}'(z_1,z_2)},
\end{split}
\end{align}
where
\begin{align}
\Delta_{F_0}'(z_1,z_2)=1-2(z_1+z_2)+(z_1-z_2)^2.
\end{align}
In this example, the mirror map \eqref{mirror_map} becomes trivial, i.e. $\log q_i=\log z_i$, $i=1,2$, and we see
that the factorization into the three-point Yukawa couplings \eqref{factorize:yukawa_4} reproduces the previous results 
in \cite{Klemm:2007in} appropriately.

\section{Conclusions}\label{sec:conclusions}

In this paper, we have thoroughly investigated the relationship between the topological B-model Yukawa couplings 
for backgrounds with non-compact directions and the exact localization formula for the A-twisted correlation 
functions of the two dimensional $\mathcal{N}=(2,2)$ gauged linear sigma models. 
Starting from the exact results for the A-twisted correlators of $\mathcal{N}=(2,2)$ gauged 
linear sigma models \cite{Closset:2015rna}, we demonstrated how to extract appropriately the B-model 
Yukawa couplings for the local nef toric Calabi-Yau varieties. First we explained that the inclusion of 
the twisted masses for the chiral matter multiplets is indispensable 
in conducting explicit calculations 
for the Jeffrey-Kirwan residue formalism. Although it has been argued in the literature that a twisted 
mass deformation is required to deal with backgrounds with non-compact directions, a comprehensive 
study has not been conducted before. We addressed this important issue and proposed an algorithm to
compute the GLSM correlation functions for local toric Calabi-Yau varieties  appropriately.

We have also checked that our prescription for the twisted mass deformations of the GLSM correlation functions 
is totally consistent with known results for the Yukawa couplings evaluated from the local mirror symmetry approach. 
Moreover, we found that the ambiguities of classical intersection numbers of a certain class of local toric Calabi-Yau 
varieties argued previously are identified with the degrees of freedom of the assignment of the proper twisted mass 
parameters. In combination with the exact localization formula for the A-twisted GLSM correlation functions, our prescription 
would provide an alternative efficient formalism to compute the B-model Yukawa couplings for generic local nef toric 
Calabi-Yau varieties.

Finally we would like to comment on possible future research directions.
Throughout this paper, we have not considered models whose target spaces are non-nef varieties \cite{Forbes:2006sj,Forbes:2006ab,Forbes:2007cy}. 
It would be interesting to extend our analysis for the GLSM correlation functions and twisted mass deformations into such intriguing examples.

In our framework, the twisted masses are introduced only for the compact directions of the target space. 
It would be interesting to clarify the physical meaning of this requirement. One possible explanation is the following.
From the viewpoint of the SUSY algebra, inclusion of the twisted masses modifies the central charges of the model 
and implies the existence of additional charged BPS particles in a specific vacuum. The appearance of such extra 
massless states can be naturally interpreted as D-branes wrapping on compact directions, as discussed 
in \cite{Gopakumar:1998ii,Gopakumar:1998jq}; 
this is known to be indispensable to regularize appropriately a singularity of the 
model.


\subsection*{Acknowledgements}

We would like to thank Masazumi Honda, Masao Jinzenji, and Yutaka Yoshida for useful discussions and comments.
The work of YH was supported in part by the grant MOST-105-2119-M-007-018 and 106-2119-M-007-019 from 
the Ministry of Science and Technology of Taiwan.
The work of MM was supported by the ERC Starting Grant no. 335739 
``Quantum fields and knot homologies'' funded by the European Research Council under the European Union's 
Seventh Framework Programme and currently by the Max-Planck-Institut f\"ur Mathematik in Bonn.

\appendix

\section{Building blocks of $I$-functions from GLSM correlators}\label{sub:I_glsm}

Let us consider a GLSM which flows in the IR to a two dimensional non-linear sigma model with a Fano 
or a Calabi-Yau variety $X$ as a target space. Following \cite{Ueda:2016wfa} (see also \cite{Bonelli:2013mma}), 
here we will show that the building blocks of the Givental $I$-function \cite{Givental:1995,Givental,Coates:2001ewh} 
for $X$ can be derived from the factors \eqref{vec_formula} and \eqref{mat_formula} in the localization formula.

First, let us reparametrize $\mathbf{u}$ and $\mathbf{d}$ as
\begin{align}
\mathbf{u}=\mathbf{x}+\frac{\mathbf{d}}{2}\hbar-\mathfrak{q}'\hbar,\qquad
\mathbf{d}=\mathfrak{q}+\mathfrak{q}'.
\end{align}
Then the factor \eqref{vec_formula} can be decomposed as
\begin{align}
Z_{\mathbf{d}}^{\textrm{vec}}(\mathbf{u};\hbar)=
I_{\textrm{pert}}^{\textrm{vec}}(\mathbf{x})\,
I_{\mathfrak{q}}^{\textrm{vec}}(\mathbf{x};\hbar)\,
I_{\mathfrak{q}'}^{\textrm{vec}}(\mathbf{x};-\hbar),
\end{align}
where
\begin{align}
I_{\textrm{pert}}^{\textrm{vec}}(\mathbf{x})&=
\prod_{\alpha\in \Delta_+}(-1)\,\alpha(\mathbf{x})^2,
\\
I_{\mathfrak{q}}^{\textrm{vec}}(\mathbf{x};\hbar)&=
\prod_{\alpha\in \Delta_+}(-1)^{\alpha(\mathfrak{q})}\,
\frac{\alpha(\mathbf{x})+\alpha(\mathfrak{q})\hbar}{\alpha(\mathbf{x})}.
\label{build_vec}
\end{align}
The factor \eqref{mat_formula} for $\Phi_a=\Phi$ with representation $R_a=R$, $R$-charge $r_a=0$ and twisted mass 
$\lambda_a=\lambda$ can be decomposed as
\begin{align}
Z_{\mathbf{d}}^{\Phi}(\mathbf{u};\hbar)&=
\begin{cases}
\prod_{\rho \in R}\prod_{p=0}^{\rho(\mathbf{d})}\left(\rho(\mathbf{u})+\lambda+p\hbar-\frac{\rho(\mathbf{d})}{2}\hbar\right)^{-1},
&
\textrm{if}\ \ \rho(\mathbf{d})\ge 0,
\\
\prod_{\rho \in R}\prod_{p=1}^{-\rho(\mathbf{d})-1}\left(\rho(\mathbf{u})+\lambda-p\hbar-\frac{\rho(\mathbf{d})}{2}\hbar\right),
&
\textrm{if}\ \ \rho(\mathbf{d})\le -1,
\end{cases}
\label{phi_c_loop}
\\
&=
I_{\textrm{pert}}^{\Phi}(\mathbf{x},\lambda)\,
I_{\mathfrak{q}}^{\Phi}(\mathbf{x},\lambda;\hbar)\,
I_{\mathfrak{q}'}^{\Phi}(\mathbf{x},\lambda;-\hbar),
\end{align}
where
\begin{align}
I_{\textrm{pert}}^{\Phi}(\mathbf{x},\lambda)&=
\prod_{\rho\in R}\frac{1}{\rho(\mathbf{x})+\lambda},
\\
I_{\mathfrak{q}}^{\Phi}(\mathbf{x},\lambda;\hbar)&=
\begin{cases}
\prod_{\rho\in R}\prod_{p=1}^{\rho(\mathfrak{q})}
\left(\rho(\mathbf{x})+\lambda+p\hbar\right)^{-1},\quad
&
\textrm{for}\ \ \rho(\mathfrak{q})\ge 0,
\\
\prod_{\rho\in R}\prod_{p=0}^{-\rho(\mathfrak{q})-1}
\left(\rho(\mathbf{x})+\lambda-p\hbar\right),\quad
&
\textrm{for}\ \ \rho(\mathfrak{q})\le -1.
\end{cases}
\label{build_mat0}
\end{align}
Similarly, the factor \eqref{mat_formula} for $\Phi_a=P$ with representation $R_a=R$, $R$-charge $r_a=2$, and twisted mass $\lambda_a=\lambda$ can be decomposed as
\begin{align}
Z_{\mathbf{d}}^{P}(\mathbf{u};\hbar)&=
\begin{cases}
-\prod_{\rho \in R}\prod_{p=0}^{-\rho(\mathbf{d})}
\left(\rho(\mathbf{u})+\lambda-p\hbar-\frac{\rho(\mathbf{d})}{2}\hbar\right),
&
\textrm{if}\ \ \rho(\mathbf{d})\le 0,
\\
-\prod_{\rho \in R}\prod_{p=1}^{\rho(\mathbf{d})-1}
\left(\rho(\mathbf{u})+\lambda+p\hbar-\frac{\rho(\mathbf{d})}{2}\hbar\right)^{-1},
&
\textrm{if}\ \ \rho(\mathbf{d})\ge 1,
\end{cases}
\\
&=
I_{\textrm{pert}}^{P}(\mathbf{x},\lambda)\,
I_{\mathfrak{q}}^{P}(\mathbf{x},\lambda;\hbar)\,
I_{\mathfrak{q}'}^{P}(\mathbf{x},\lambda;-\hbar)
\prod_{\rho \in R} (-1)^{\rho(\mathbf{d})},
\end{align}
where
\begin{align}
I_{\textrm{pert}}^{P}(\mathbf{x},\lambda)&=
-\prod_{\rho\in R}\left(\rho(\mathbf{x})+\lambda\right),
\\
I_{\mathfrak{q}}^{P}(\mathbf{x},\lambda;\hbar)&=
\begin{cases}
\prod_{\rho\in R}\prod_{p=1}^{-\rho(\mathfrak{q})}
\left(-\rho(\mathbf{x})-\lambda+p\hbar\right),\quad
&
\textrm{for}\ \ \rho(\mathfrak{q})\le 0,
\\
\prod_{\rho\in R}\prod_{p=0}^{\rho(\mathfrak{q})-1}
\left(-\rho(\mathbf{x})-\lambda+p\hbar\right)^{-1},\quad
&
\textrm{for}\ \ \rho(\mathfrak{q})\ge 1.
\end{cases}
\label{build_mat2}
\end{align}

After taking the above decomposition, the ingredients $I_{\mathfrak{q}}^{\textrm{vec}}(\mathbf{x};\hbar)$ in \eqref{build_vec}, 
$I_{\mathfrak{q}}^{\Phi}(\mathbf{x},\lambda;\hbar)$ in \eqref{build_mat0} and $I_{\mathfrak{q}}^{P}(\mathbf{x},\lambda;\hbar)$ 
in \eqref{build_mat2} are known to provide the building blocks of 
the Givental $I$-function for the variety $X$. There $\mathbf{x}$ and $\lambda$ 
are identified with the equivariant cohomology elements or the Chern roots of $X$ and the equivariant parameter acting on $X$, respectively 
\cite{Bonelli:2013mma,Ueda:2016wfa,Inoue16,Kim16,Gerhardus:2018zwb}. Starting from the $I$-function, one can find an associated 
quantum differential equation called the Picard-Fuchs equation from which the mirror map and genus zero Gromov-Witten invariants of $X$ 
can be evaluated \cite{Givental:1995,Givental,Coates:2001ewh}.

\section{Local B-model Yukawa couplings of local $dP_2$}\label{app:dp2}

The exact B-model Yukawa couplings $Y_{z_iz_jz_k}^{(0)}(\mathbf{z})=Y_{z_iz_jz_k}(z_1,z_2,z_3)\big|_{p=q=0}$ for $p=q=0$ of the local del Pezzo surface $K_{dP_2}$ 
defined in \eqref{loc_dp2_correl} have the following expressions:
\small
\begin{align}
\begin{split}
Y_{z_1z_1z_1}^{(0)}(\mathbf{z})&=
\big(4 z_1^2 \left(z_2-z_3\right)^2 \left(z_2+z_3-2\right)
+z_1 \left(z_2-1\right) \left(z_3-1\right) \left(5 z_3-z_2 \left(9 z_3-5\right)-1\right)\big)
/\Delta_{dP_2}(\mathbf{z}),
\\
Y_{z_1z_1z_2}^{(0)}(\mathbf{z})&=
\big(-4 \left(2 z_2^3-\left(3 z_3+2\right) z_2^2+\left(z_3+1\right)^2 z_2-z_3\right) z_1^2+\big(\left(9 z_3^2-8 z_3-2\right) z_2^2
-2\left(7 z_3^2-7 z_3-1\right) z_2
\\
&\ \ \
+4 z_3^2-4 z_3-1\big) z_1+\left(z_2-1\right) \left(z_3-1\right)\big)/2\Delta_{dP_2}(\mathbf{z}),
\\
Y_{z_1z_2z_2}^{(0)}(\mathbf{z})&=
\big(4 \left(z_2-1\right) z_2 \left(z_2-2 z_3+1\right) z_1^2
+z_2 \left(z_3-1\right) \left(8 z_3-z_2 \left(9 z_3-5\right)-4\right) z_1\big)
/\Delta_{dP_2}(\mathbf{z}),
\\
Y_{z_1z_2z_3}^{(0)}(\mathbf{z})&=
\big(4 \left(z_3 z_2^2+\left(z_3^2-4 z_3+1\right) z_2+z_3\right) z_1^2+\big(\left(9 z_3^2-14 z_3+4\right) z_2^2-2 \left(7 z_3^2-10 z_3+2\right) z_2
\\
&\ \ \
+4 z_3^2-4 z_3-1\big) z_1+\left(z_2-1\right) \left(z_3-1\right)
\big)/2\Delta_{dP_2}(\mathbf{z}),
\\
Y_{z_2z_2z_2}^{(0)}(\mathbf{z})&=
\big(
-16 \left(z_2^2-z_3^2\right) z_1^3-8 \left(4 z_2^3-4 z_3 z_2^2-\left(z_3^2+2 z_3-2\right) z_2-2 z_3^3+2 z_3^2+z_3\right) z_1^2
\\
&\ \ \
+\left(\left(45 z_3^2-52 z_3+4\right) z_2^2+\left(20 z_3^2-38 z_3+20\right) z_2-8 z_3^2+8 z_3+1\right) z_1
\\
&\ \ \
+\left(3 z_2+1\right) \left(z_3-1\right)\big)/4\Delta_{dP_2}(\mathbf{z}),
\\
Y_{z_2z_2z_3}^{(0)}(\mathbf{z})&=
\big(
16 \left(z_2^2-z_3^2\right) z_1^3+8 \left(2 z_2^3-2 z_2^2-z_3^2 z_2-2 z_3^3+2 z_3^2+z_3\right) z_1^2+\big(\left(-9 z_2^2-4 z_2+8\right) z_3^2
\\
&\ \ \
+2\left(4 z_2^2+3 z_2-4\right) z_3-1\big) z_1+\left(z_2-1\right) \left(z_3-1\right)
\big)/4\Delta_{dP_2}(\mathbf{z}),
\nonumber
\end{split}
\end{align}\normalsize
and 
\small
\begin{align}
\begin{split}
&
Y_{z_1z_1z_3}^{(0)}(z_1,z_2,z_3)=Y_{z_1z_1z_2}^{(0)}(z_1,z_3,z_2),\quad Y_{z_1z_3z_3}^{(0)}(z_1,z_2,z_3)=Y_{z_1z_2z_2}^{(0)}(z_1,z_3,z_2),\\
&
Y_{z_2z_2z_2}^{(0)}(z_1,z_2,z_3)=Y_{z_3z_3z_3}^{(0)}(z_1,z_3,z_2),\quad Y_{z_2z_2z_3}^{(0)}(z_1,z_2,z_3)=Y_{z_2z_3z_3}^{(0)}(z_1,z_3,z_2),
\nonumber
\end{split}
\end{align}\normalsize
where
\small
\begin{align}
\begin{split}
\Delta_{dP_2}(\mathbf{z})&=
16(z_2-z_3)^2z_1^3+8\left((2z_2+2z_3-3)(z_2-z_3)^2-(z_2+z_3)(1-z_2)(1-z_3)\right)z_1^2
\\
&\ \ \
+\left(4(9z_2z_3-2z_2-2z_3+2)(z_2+z_3)+1-30z_3z_2-27z_2^2z_3^2\right)z_1-(1-z_2)(1-z_3).
\nonumber
\end{split}
\end{align}\normalsize
These expressions indeed agree with the local B-model Yukawa couplings in \cite{Forbes:2005xt} with $x=1/4$ and $y=-1/4$.

\section{B-model Yukawa couplings of local $A_2$ geometry}\label{app:a2}

The exact B-model Yukawa couplings of the local $A_2$ geometry 
defined in \eqref{loc_A2_correl} have the following expressions:
\small
\begin{align}
\begin{split}
Y_{z_1z_1z_1}(\mathbf{z})&=
2 z_1 \left(-16 z_1^2 z_2^2 (3 z_3 (9 z_3 (3 z_2 (6 z_3-1)+4 z_3-5)+14)-4)
\right.
\\
&\ \ \
+4 z_1 \left(324 (z_2-2) z_2^2 z_3^3
+2 (3 z_2 (32-9 (z_2-3) z_2)-8) z_3^2+(8-92 z_2) z_3+11 z_2-1\right)
\\
&\ \ \
\left.
+(z_2 (6 z_3-1)-4 z_3+1) (z_2 (9 z_3 (3 z_2 z_3-2)+4)+4 z_3-1)\right)
/3 (1-4 z_1)^2\Delta_{A_2}(\mathbf{z}),
\\
Y_{z_1z_1z_2}(\mathbf{z})&=-
4 z_1 \left(-108 z_2^2 z_3^3 (4 z_1 (3 z_2+1)-3 z_2+3)+z_3^2 (9 z_2 (3 z_2 (4 z_1 (2 z_2+5)-2 z_2-1)+16)-16)\right.
\\
&\ \ \
\left.
-2 z_2 z_3 (21 (4 z_1-1) z_2+34)+4 z_2 ((4 z_1-1) z_2+2)+8 z_3-1\right)
/3 (1-4 z_1)\Delta_{A_2}(\mathbf{z}),
\\
Y_{z_1z_1z_3}(\mathbf{z})&=-
2 z_1 \left(-54 z_2^2 z_3^3 (z_1 (60 z_2+8)-15 z_2+6)+z_3^2 (9 z_2 (3 z_2 (4 z_1 (8 z_2+11)-8 z_2-7)+16)-16)\right.
\\
&\ \ \
\left.
-4 z_2 z_3 (33 (4 z_1-1) z_2+17)+8 z_2 ((8 z_1-2) z_2+1)+8 z_3-1\right)
/3 (1-4 z_1)\Delta_{A_2}(\mathbf{z}),
\\
Y_{z_1z_2z_2}(\mathbf{z})&=
2 \left(-54 z_2^2 z_3^3 (4 z_1 (3 z_2+2)-3 z_2+2)+z_3^2 (3 z_2 (9 z_2 (4 z_1 (z_2+5)-z_2-3)+32)-16)\right.
\\
&\ \ \
\left.
-2 z_2 z_3 (21 (4 z_1-1) z_2+22)+z_2 (4 (4 z_1-1) z_2+5)+8 z_3-1\right)
/3\Delta_{A_2}(\mathbf{z}),
\\
Y_{z_1z_2z_3}(\mathbf{z})&=
\left(-27 z_2^2 z_3^3 (4 z_1 (15 z_2+4)-15 z_2+4)+z_3^2 (3 z_2 (9 z_2 (4 z_1 (4 z_2+11)-4 z_2-9)+44)-16)\right.
\\
&\ \ \
\left.
+z_2 z_3 (132 (1-4 z_1) z_2-65)+8 z_2 ((8 z_1-2) z_2+1)+8 z_3-1\right)
/3\Delta_{A_2}(\mathbf{z}),
\\
Y_{z_1z_3z_3}(\mathbf{z})&=
2 \left(-54 (4 z_1-1) z_2^3 z_3^2 (3 z_3-2)-z_2^2 (4 z_1 (3 z_3 (9 (z_3-5) z_3+32)-16)\right.
\\
&\ \ \
\left.
+3 z_3 (9 z_3 (z_3+3)-32)+16)+z_2 \left(42 z_3^2-44 z_3+8\right)+z_3 (5-4 z_3)-1\right)
/3\Delta_{A_2}(\mathbf{z}),
\\
Y_{z_2z_2z_2}(\mathbf{z})&=
4 \left(-(4 z_1-1) z_2^2 (3 z_3 (3 z_3-2) (12 z_3-7)-4)+z_2 (4 z_3 (12 z_3-5)+2)-(1-4 z_3)^2\right)
/3\Delta_{A_2}(\mathbf{z}),
\\
Y_{z_2z_2z_3}(\mathbf{z})&=
2 \left(-(4 z_1-1) z_2^2 (3 z_3 (9 z_3 (4 z_3-11)+44)-16)\right.
\\
&\ \ \
\left.
+2 z_2 (z_3 (60 z_3-31)+4)-(1-4 z_3)^2\right)
/3\Delta_{A_2}(\mathbf{z}),
\\
Y_{z_2z_3z_3}(\mathbf{z})&=
4 \left(-(4 z_1-1) z_2^2 (3 z_3 (9 (z_3-5) z_3+32)-16)
\right.
\\
&\ \ \
\left.
+z_2 \left(48 z_3^2-44 z_3+8\right)+z_3 (5-4 z_3)-1\right)
/3\Delta_{A_2}(\mathbf{z}),
\\
Y_{z_3z_3z_3}(\mathbf{z})&=
\left(
-2 (4 z_1-1) z_2^2 (3 z_3 (9 (z_3-8) z_3+80)-64)
\right.
\\
&\ \ \
\left.
+4 z_2 (z_3 (33 z_3-52)+16)-8 z_3^2+22 z_3-8\right)
/3\Delta_{A_2}(\mathbf{z}),
\nonumber
\end{split}
\end{align}\normalsize
where
\small
\begin{align}
\begin{split}
\Delta_{A_2}(\mathbf{z})&=
729 (1-4 z_1)^2 z_2^4 z_3^4+108 (4 z_1-1) z_2^3 (9 z_3-2) z_3^2+2 z_2^2 (4 z_1 (9 z_3 (3 z_3 (4 z_3-7)+8)-8)
\\
&\ \ \
+9 z_3 (3 z_3 (4 z_3+5)-8)+8)-4 z_2 (4 z_3-1) (9 z_3-2)+(1-4 z_3)^2.
\nonumber
\end{split}
\end{align}\normalsize

\subsubsection*{Derivation of $\Delta_{A_2}(\mathbf{z})$}

The factor $\Delta_{A_2}(\mathbf{z})$ appearing in the above 
expressions can be regarded as a discriminant which describes the degenerate points of the mirror 
curve \cite{Hori:2000kt,Hori:2000ck}. To find a mirror curve explicitly, let us first consider the mirror local 
Calabi-Yau threefold of the local $A_2$ geometry defined by
\begin{align}
\begin{split}
X^{\vee}=\bigg\{
(\omega_+, \omega_-, x_1,x_2,x_3,x_4,x_5,x_6) \in {\IC}^2 \times 
\left({\IC}^*\right)^6\, \bigg|\, 
&
\omega_+\omega_-=\sum_{i=1}^6x_i,\
x_1x_2=z_1 x_3^2
\\
&\ \ \
x_3x_5=z_2x_4^2,\ x_4x_6=z_3x_5^2\bigg\},
\label{mirror_cy}
\end{split}
\end{align}
where each $x_i$, $i=1,\ldots, 6$ is the mirror coordinate corresponding to $X_i$ in Figure \ref{fig:local_A2}. Let us take a 
local coordinate $x_1=y$, $x_3=0$, $x_4=x$, which corresponds to the mirror of a brane wrapping on a Lagrangian submanifold located at 
the external leg of the local atlas around $X_1=X_3=X_4=0$. Then $x$ becomes the open string moduli parameter in the B-model. 
From \eqref{mirror_cy}, we obtain a mirror curve in $X^{\vee}$ at $\omega_+=0$ or $\omega_+=0$ as
\begin{align}
y^2+\left(1+x+z_2x^2+z_2^2z_3x^3\right)y+z_1=0.
\label{mirror_curve}
\end{align}
The branch points can be obtained from
$$
\left(1+x+z_2x^2+z_2^2z_3x^3\right)^2-4z_1=0,
$$
and the discriminant for $x$ of the above equation given by
$$
2^{12} z_1^3 z_2^{16}z_3^6\,
\Delta_{A_2}(\mathbf{z})
$$
describes the degenerate points of the mirror curve \eqref{mirror_curve}.


\end{document}